\documentclass[twocolumn]{aa}  
\usepackage{graphicx}
\usepackage{txfonts}
\usepackage[T1]{fontenc}
\usepackage{graphicx}	
\usepackage{amsmath}	
\usepackage{amssymb}	
\usepackage{times}
\usepackage{longtable}

\begin{document}

\title{Outflows from GRB hosts are ubiquitous: Kinematics of $z<0.3$ GRB-SN hosts resolved with FLAMES  \thanks{based on ESO proposal 092.D-0389, PI C. Th\"one} } 

\titlerunning{Kinematics of $z<0.3$ GRB-SN hosts resolved with FLAMES}

\author{C. C. Th\"one \inst{1} \and L. Izzo  \inst{1,3} \and H. Flores \inst{2} \and A. de Ugarte Postigo \inst{1,3} \and S. D. Vergani \inst{2}  \and J. F. Ag\"u\'i Fern\'andez\inst{1} \and D. A. Kann\inst{1} \and L. Christensen \inst{3} \and S. Covino \inst{4} \and M. Della Valle \inst{5} \and F. Hammer \inst{2} \and A. Melandri \inst{4} \and M. Puech \inst{2} \and M. A. Rodrigues \inst{6} \and J. Gorosabel \inst{1,7,8}\textdagger
}

\institute{
Instituto de Astrof\'isica de Andaluc\'ia - CSIC, Glorieta de la Astronom\'ia s/n, 18008 Granada, Spain\\
\email{cthoene@iaa.es} \and
GEPI, Observatoire de Paris, PSL University, CNRS, 5 Place Jules Janssen, 92190 Meudon, France \and
Dark Cosmology Centre, Niels-Bohr-Institute, Univ. of Copenhagen, Jagtvej 128, 2200 Copenhagen \and
INAF, Osservatorio Astronomico di Brera, Via Bianchi 46, 23807 Merate (LC), Italy \and
Astronomical Observatory of Capodimonte in Naples (OACN), Salita Moiariello, Napoli, 80131, Italy \and
University of Oxford, Department of Physics, Keble Road, Oxford OX1 3RH, UK \and
Departamento de F\'isica Aplicada I, E.T.S. Ingenier\'ia, Universidad del Pa\'is-Vasco UPV/EHU, Alameda de Urquijo s/n, 48013 Bilbao, Spain \and
Ikerbasque, Basque Foundation for Science, Alameda de Urquijo 36-5, 48008 Bilbao, Spain
}

\date{Received ; accepted }

\abstract{The hosts of long duration gamma-ray bursts (GRBs) are predominantly starburst galaxies at subsolar metallicity. At redshifts $z<1$, this implies that most of them are low-mass galaxies similar to the populations of blue compact dwarfs and dwarf irregulars. What triggers the massive star-formation needed for producing a GRB progenitor is still largely unknown, as are the resolved gas properties and kinematics of these galaxies and their formation history. Here we present a sample of six spatially resolved GRB hosts at $z<0.3$ observed with 3D spectroscopy at high spectral resolution (R\,$=$\,8,000-13,000) using FLAMES/VLT. We analyzed the resolved gas kinematics of the full sample and the abundances in a subsample with strong enough emission lines. Only two galaxies show a regular disk-like rotation field, another two are dispersion-dominated, and the remaining ones have two narrow emission components associated with different parts of the galaxy but no regular rotation field, which might indicate a recent merger. All galaxies show evidence for broad components underlying the main emission peak with $\sigma$ of 50--110 km\,s$^{-1}$. This broad component is more metal-rich than the narrow components, it is blueshifted in most cases, and it follows a different velocity structure. We find a weak correlation between the star-formation rate and the width of the broad component, its flux compared to the narrow component, and the maximum outflow velocity of the gas, but we do not find any correlation with the star-formation density, metallicity or stellar mass. We hence associate this broad component with a metal-rich outflow from star-forming regions in the host. The GRB is not located in the brightest region of the host, but is always associated with some star-forming region showing a clear wind component. Our study shows the great potential of 3D spectroscopy to study the star-formation processes and history in galaxies hosting extreme transients, the need for high signal-to-noise (S/N), and the perils using unresolved or only partially resolved data for these kinds of studies.}


\maketitle



\section{Introduction}
Long gamma-ray burst (LGRB) progenitors have indubitably been identified as massive stars through their connection to broad-line Type Ic  supernovae (SNe) coincident with the GRB \citep[for a recent review see][]{CanoRev}. Models suggest that their progenitors are Wolf-Rayet (WR) stars, massive stars stripped of their H and He envelopes, with low metallicity where stellar winds are weaker, and they retain enough angular momentum necessary for the GRB jet to form \citep{WoosleyHeger06}. Massive star-formation in low metallicity gas at low redshift happens primarily in dwarf starburst galaxies. Unsurprisingly, dwarf galaxies dominate the population of low redshift GRB hosts: The average luminosity and stellar masses of GRB hosts are --19\,mag or log~M*\,$=$\,9.0 M$_\odot$ at $z\approx 0$, but they rise to log~M*\,$=$\,9.6\,M$_\odot$ and $>$10\,M$_\odot$ at redshifts of $1<z<2$ and $z>2$ \citep{PerleySHOALS2, Palmerio19}. GRB hosts at low redshift are similar to the galaxy populations of blue compact dwarfs (BCDs, defined as M$_\mathrm{abs}<$~--18 mag and size $<1$\,kpc), and dwarf irregulars (dIrrs); a few have also been found to be dwarf spiral galaxies (e.g., GRB 980425 and GRB 060505). A few exceptions are GRB 171205A, the third closest GRB ever detected and hosted by a large spiral with a mass of log~M*$\sim$10.1 (\citealt{PerleyGCN, Izzo19}; Th\"one et al. in prep.), the even larger spiral host of GRB 190829A (Izzo et al. in prep.) and the more distant face-on grand design spiral host of GRB 990705 \citep{LeFloch02,Hunt14}.

Since GRBs are distant and occur in small galaxies rarely monitored by high angular-resolution surveys, it is unlikely for the forseable future to image the progenitor of a GRB, hence we need to infer properties of the progenitor star from its environment. The average redshift of LGRBs is $z\sim 2.2$ \citep{Coward13} where 1\,arcsec corresponds to a physical size of 8\,kpc, preventing spatially resolved observations with current facilities. To date only a handful of GRB hosts have been studied with IFU data: GRB 980425 \citep{Christensen08, Kruehler17}, the SN-less long GRB 060505 \citep{Thoene08, Thoene14}, GRB 100316D \citep{Izzo17a}, which contains one of the datasets presented in this paper, and GRB 111005A, another potential SN-less GRB \citep{Tanga17, Michalowski18}. 

The origin of star-formation in dwarf galaxies is still largely unknown. Their star-burst episodes seem to only last a few tens of Myr followed by Gyrs of quiescence \citep{Lee09,McQuinn10,Zhao11}. The small potential well allows to disrupt molecular clouds with only a few SN explosions and quench SF. Some starbursts might be triggered by interactions \citep[see e.g.,][]{Bekki08, vanZee98}, suggestions are up to $>$60\% \citep{PerezGallego11}, but many galaxies seem to be isolated (see e.g. the SIGRID sample, \citealt{NichollsSIGRID}, or the LITTLE THINGS survey, \citealt{Hunter12}). However, low luminosity neighbors may easily go undetected and only show up in HI gas \citep[see e.g.][]{Ashley13, Ashley17}. Other possible mechanisms include inflows of gas  \citep{Elmegreen12, Verbeke14}, or stellar feedback and ram-stripping, removing gas that is later re-accreted \citep{Ashley17}. Very likely not one single mechanism can explain all starburst activity in dwarf galaxies \citep[see e.g.][]{Koleva14}.

GRB hosts are only rarely found to be interacting systems (e.g. GRB 090323, \citealt{Savaglio12}, GRB 090426, \citealt{Thoene11}, GRB 120422A, \citealt{Schulze14}, GRB 080810, \citealt{Wiseman17}, and possibly GRB 060418 and GRB 050820, \citealt{Chen12}). 
Inflows have been suggested as a SF trigger from resolved HI and radio continuum maps \citep{Michalowski12, Michalowski15}. HI might also trace past interactions \citealt[][]{}[e.g. for the host of GRB\,980425][]{Arabsalmani19}). Even in the most compact dwarfs, kinematics of hot gas, HI gas and stellar kinematics do not necessarily trace each other, supporting the interaction scenario for star-formation  \citep{Johnson12, Koleva14,Ashley17}. 

Star-burst dwarf galaxies often show broad components in nebular emission lines with velocities up to a thousand km\,s$^{-1}$ \citep{Izotov07, Telles14}. Possible explanations are 1) SN explosions or stellar winds form large bubbles \citep{Telles14}, 2) turbulent mixing layers on the surface of dense gas clouds \citep{Westmoquette07, James09} or 3) AGN activity, although the latter has been mostly ruled out \citep{Izotov07, James09}. Green-pea (GP) galaxies \citep{Cardamone09}, extreme BCDs in an early starburst phase characterized by strong [OIII] emission, often  show broad components associated with SN driven winds \citep{AmorinGPkin}. The frequency of these components in the starburst galaxy population is still a matter of debate \citep[see e.g.][]{James10}. 

Galactic winds are linked to starburst activity. At low redshift they can be directly observed (e.g. in M82) while at high redshifts they are studied via absorption lines \citep[for reviews see][]{Veilleux05, Rupke18}. Emission and absorption lines probe different parts of the winds, but beyond the local Universe, observing both has rarely been done \citep{Erb12, Wood15}. Recently, the MEGAFLOW sample began to study  galactic winds detected in background QSO absorbers and their galaxy counterparts in emission with MUSE \citep{Schroetter16, Zabl19}. Galactic winds are now seen as a crucial factor in explaining and constraining the shape of the mass-metallicity relation \citep{Mannucci10,Chisholm18} and the enrichment of the intergalactic medium (IGM).


In this paper we present the first sample of long GRB hosts observed at high spatial and spectral resolution with FLAMES/VLT which is complete up to $z=0.3$ for GRBs discovered until early 2013 and observable from the VLT. This comprises the hosts of GRB 020903, GRB 030329, GRB 031203, GRB 050826, GRB 060218, GRB 100316D and GRB 120422A. For GRB 120422A, we did not obtain data due to an incorrect pointing of the instrument. In Sect. 2 we present the observations and analysis, Sect. 3 details the result for the different hosts regarding kinematics and abundances and Sect. 4 discusses the results. Throughout the paper we use a flat lambda CDM cosmology as constrained by Planck with $\Omega_m=0.315$, $\Omega_\Lambda=0.685$ and $H_0=67.4$ \citep{Planck18}. 

\section{Observations and analysis}
\begin{figure*}
\begin{center}
	\includegraphics[width=18.5cm]{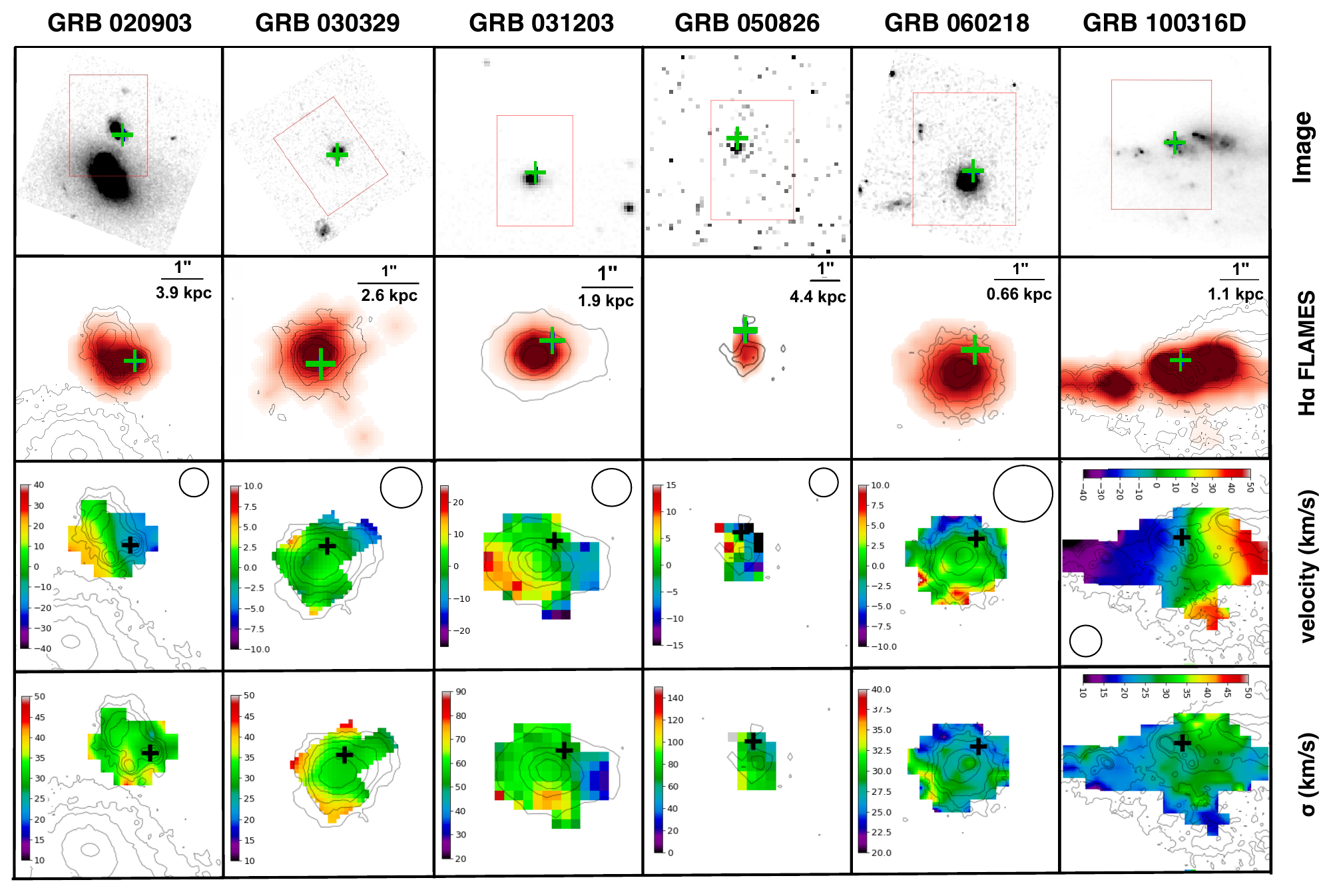} 
    \caption{GRB hosts observed with FLAMES. {\it Top row:} Imaging of the field, the position of the GRB is marked by a green cross, the FLAMES FOV is shown by a red rectangle. Images are from HST for GRB 020903, GRB 030329, GRB 060218 and GRB 100316D \citep{Fruchter06,Svensson10}. For GRB 050826 we used data from the PanSTARRS survey, for GRB 031203 imaging from FORS 2/VLT from the ESO archive. {\it Second row:} H$\alpha$ maps interpolated to $\sim$0.02'' for better visualization with contours overplotted using the broad-band image in the top row. {\it Third and fourth row}: Velocity and dispersion maps of the sample derived from a single Gaussian fit to H$\alpha$. The maps are interpolated to the resolution of the image used for the contour plot (0.02'' for the ones with HST imaging, 0.25'' for the FORS and PanSTARRS images of the hosts of GRB 031203 and GRB 050826). The nominal resolution including the seeing is shown by a circle for each host. In all plots north is up and east is left.}
    \label{Fig:sample}
    \end{center}
\end{figure*}

\begin{figure*}
	\centering
	\includegraphics[width=17cm]{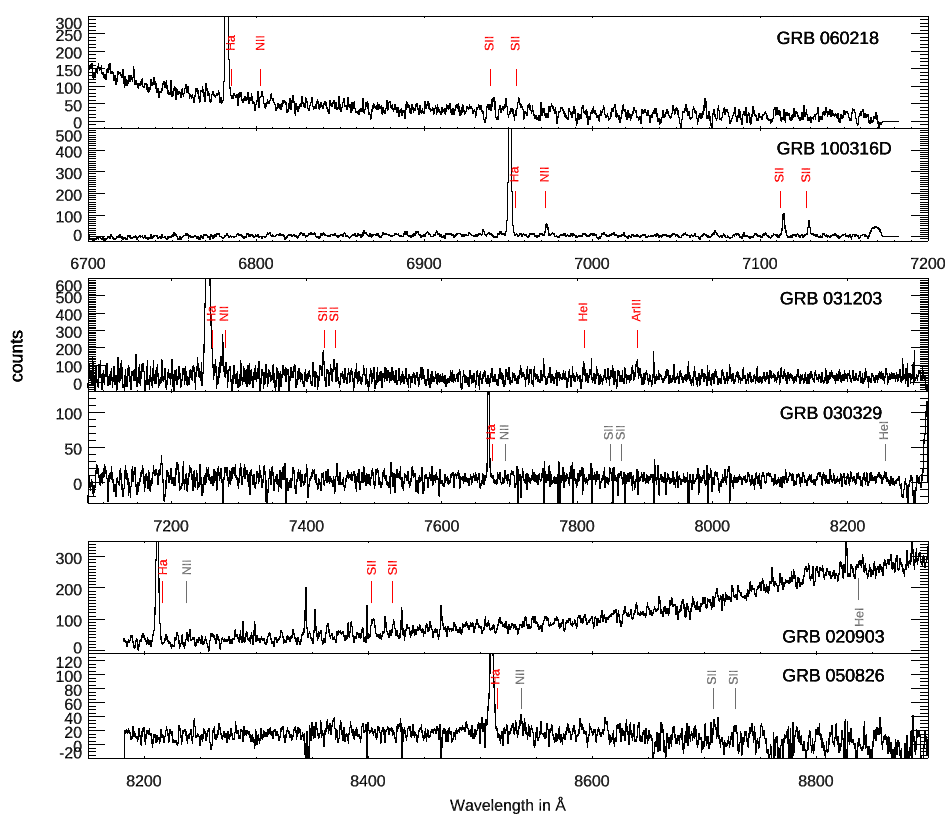}
    \caption{Integrated spectra of the 6 galaxies in our sample, grouped by the different grating settings used (top to bottom: LR 6, 7 and 8). Red lines are detected transitions, gray tick marks indicate the position of emission lines that are not detected in that particular host. For each spectrum we summed the flux for all regions where H$\alpha$ is detected and subtracted the average sky background using at least 9 spaxels outside the host galaxy. The final spectra have been smoothed with a Gaussian kernel of 5.}
    \label{fig:integratedspecs}
\end{figure*}

The GRB host sample was observed with FLAMES/GIRAFFE at the VLT between Nov. 2013 and March 2014 with a total observing time of 22\,h. Observations were done in ARGUS mode with the 0.3'' sampling size, which provides a 6.6'' $\times$ 4.2'' field-of-view (FOV, see Fig. \ref{Fig:sample}). We used the low resolution grating at three different incidence angles called LR6, 7 and 8, resulting in spectral resolutions of R\,$=$\,13,500, 8200 and 10,000 respectively. The aim was to cover at least the range from H$\alpha$ to [SII] at the different redshifts of the galaxies. The exact wavelength coverage for each spectrum is shown in Fig. \ref{fig:integratedspecs}.

The nominal seeing varies as observations were performed at different dates. The values are 0.7'' for GRB 020903 and GRB 030329, 0.8'' for GRB 031203 and GRB 100316D, 1.0'' for GRB 050826 and 1.2'' for GRB 060218. The ARGUS data reduction was done using the standard ESO pipeline (version 2.11) without the sky subtraction option. To verify the fiber-to-fiber wavelength calibration, we controlled the wavelength of two skylines in the data cube. No absolute flux calibration was performed.

The sample was analyzed with dedicated software in IDL and IRAF, partially based on tools presented in \cite{Flores06,Yang08}. For the emission line maps we integrated the flux over the emission line and subtracted the continuum from a line-free region around the emission line. This approach was chosen since the emission lines have shapes often deviating from a pure Gaussian. To obtain the velocity maps we fit H$\alpha$ with a single Gaussian, even if several components are present, hence this shows the properties of the dominant line component. Using several sky emission lines we derive an instrumental resolution of FWHM$=$39.9 km\,s$^{-1}$ ($\sigma$ of 16.9 km\,s$^{-1}$). The dispersion $\sigma$ is corrected for the instrumental resolution by subtracting it in quadrature. All maps are interpolated to $\sim$0.02'' for better visualization. Multicomponent fits to individual emission lines for kinematical analysis were done with {\tt ngaussfit} in IRAF and PAN in IDL \citep[Peak ANalysis,][]{PAN1, Westmoquette07}. The final maps are plotted in Fig. \ref{Fig:sample}.

Metallicities were obtained using the N2-parameter [\ion{N}{II}]$\lambda$6585/H$\alpha$ taking the recalibration from the CALIFA sample \citep{MarinoZ}. Relative line fluxes for [NII] and H$\alpha$ were derived from a single Gaussian fit, due to their proximity in wavelength, an absolute flux calibration is not needed to derive the ratio of the two lines.

\section{Results of the individual hosts}\label{sect:results}
In Fig. \ref{fig:integratedspecs} we plot the integrated spectra of the galaxies in the sample. H$\alpha$ is detected in all galaxies while the detection of further lines depends on the S/N of the spectra. Only for two hosts, GRB 031203 and GRB 100316D, are we able to obtain 2D maps of emission line ratios and abundances. In the following we give a brief overview of literature results for each GRB host and the detected emission lines after which we focus on the kinematic analysis and the resolved abundances for those two hosts where this has been possible.

\begin{table*}
\caption{GRB host sample global properties. }
\label{tab:global}
\centering \small     
\begin{tabular}{l l l l l l l l l l l l }       
\hline\hline                 
GRB			&$z $		& log M*		& M$_B$	& r$_\mathrm{80}$ &  r$_\mathrm{50}$ & d$_\mathrm{GRB}$	&SFR	& SSFR& $\Sigma$SFR		& \multicolumn{2}{c}{12+log(O/H)}	\\  
			&		&\scriptsize{M$_\odot$}&\scriptsize{mag}   	&\scriptsize{kpc}&\scriptsize{kpc}&\scriptsize{kpc}&\scriptsize{M$_\odot$yr$^{-1}$}	&\scriptsize{Gyr$^{-1}$}&\scriptsize{M$_\odot$yr$^{-1}$kpc$^{-2}$}&	\scriptsize{this work}		& \scriptsize{lit.} \\ 
\hline                       
020903		&0.251	&8.94		&--19.34	&1.43$^{e}$ 	&0.91		& 1.1		&0.45  		 &0.92	& 0.070  	&$<$8.1	& 7.98--8.07$^{a}$		\\ 
030329		&0.169	&7.70		&--16.52	&1.03$^{e}$ 	&0.54$^{i}$		& 1.0		&0.14   		&4.74	& 0.042	&$<$8.0	& 7.7--8.0$^{b,c,d}$	\\ 
031203		&0.106	&8.86		&--18.52	&1.79	&1.04$^{i}$		& 1.1		&4.3			&2.48	& 0.094 	&8.1		& 8.1$^{c}$			\\ 
050826		&0.297	& 9.99		&--20.28	&6.21	&3.88		& 3.1		&1.39$^{e}$	&0.17 	&0.03	&$<$8.45 	& 8.8$^{a}$			\\ 
060218		&0.033	& 7.40		&--15.92	&0.55$^{e}$	&0.37$^{i}$		& 0.5		&0.05		&1.82	& 0.053	&$<$7.88 & 7.6$^{f}$			\\ 
100316D		&0.059	&9.39	&--18.8$^{d}$	&3.96&2.55$^{i}$	 	& 0.6		&1.2$^{h}$	&1.41   & 0.024	&8.25 	& 8.0--8.2$^{g,h}$		\\ 
\hline                                
\end{tabular}
\tablefoot{Stellar masses have been obtained by fitting CIGALE \citep{CIGALE} SED models to available photometry from the literature (see Appendix). B-band luminosities are from \cite{Svensson10} based on literature photometry (see that paper for details). For GRB 031203, GRB 050826 and GRB 100316D we derived r$_\mathrm{80}$, for GRB\,020903 and GRB\,050826 r$_\mathrm{50}$ from FORS2, PanSTARRS and HST imaging used for the contours in Fig.\ref{Fig:sample}, the remaining values are from \cite{Svensson10} and \cite{Japelj18}. d$_\mathrm{GRB}$ is the distance of the GRB location from the spaxel containing the brightest H$\alpha$ emission. SFRs are based on UV luminosities and taken from \cite{Michalowski15} when not indicated differently, SSFRs are the SFR divided by the stellar mass, SFR densities are calculated as $\Sigma$SFR=SFR/$(\pi\,r_\mathrm{80}^2)$. Metallicities are from the global host spectra using the N2 parameter and the \cite{MarinoZ} calibration.  \\
{\bf References.} $^{(a)}$\,\citet{Levesque10b}, $^{(b)}$\,\citet{Thoene07}, $^{(c)}$\,\citet{Levesque10a}, $^{(d)}$\,\citet{Starling12}, $^{(e)}$\,\citet{Svensson10}, $^{(f)}$\,\citet{Wiersema07}, $^{(g)}$\,\citet{Levesque11},
$^{(h)}$\,\citet{Izzo17a},
$^{(i)}$\,\citet{Japelj18}.}
\end{table*}

\begin{table*}
\caption{GRB host sample global kinematical properties.}
\label{tab:globalkin}
\centering       
\begin{tabular}{l l l l l l l l l}       
\hline\hline                 
GRB			& $\Delta$v	& $\sigma_\mathrm{int}$ & $\sigma_0$ & log M$_\mathrm{dyn,rot}$ & log M$_\mathrm{dyn,\sigma}$& log M*& $v_\mathrm{esc}$ \\  
		         &\scriptsize{km\,s$^{-1}$}&\scriptsize{km\,s$^{-1}$}&\scriptsize{km\,s$^{-1}$}&\scriptsize{M$_\odot$}&\scriptsize{M$_\odot$}&\scriptsize{M$_\odot$}&\scriptsize{km\,s$^{-1}$}\\ 
\hline                       
020903		&59			&34		&28.4& 8.54&8.96&8.94&78\\ 
030329		&15			&27		 &27.9&7.09 &8.72&7.70&70\\ 
031203		&53			&39		&11.7& 8.51&8.61&8.86&47\\ 
050826		&43			&55  	 &71.6& 8.90&10.3&9.99&172\\ 
060218		&23			&24	 	&22.2&7.49 &8.43&7.40&68\\ 
100316D		&92		    &31	 	&19.4& 9.59&9.19&9.39&61\\ 
\hline                                
\end{tabular}
\tablefoot{$\Delta$v are the difference between minimum and maximum velocity in the 2D velocity map, v$_\mathrm{rot}$=0.5$\Delta$v is also called $v_\mathrm{shear}$ (see Sect. \ref{sect:discussion}). $\sigma_\mathrm{int}$ is derived from a single Gaussian fit to the integrated spectra of each galaxy. $\sigma_0$ is the flux weighted dispersion of H$\alpha$ of all spaxels with sufficient S/N in H$\alpha$ \citep[see][]{HerenzLARS}. The FOV of the host of GRB 100316D does not comprise the entire galaxy. Dynamical masses are derived from $\Delta$v, $\sigma_0$, r$_\mathrm{80}$ and r$_\mathrm{50}$ (see Tab. \ref{tab:global}), the escape velocity of the galaxy is derived from the dynamical mass used for the respective systems. For the calculations of M$_\mathrm{dyn,rot}$, M$_\mathrm{dyn,\sigma}$ and $v_\mathrm{esc}$ see Sect. \ref{sect:discussion}). }
\end{table*}

\begin{table*}
\caption{Results of the multicomponent fits for the entire galaxy  and different integrated regions in some of the galaxies.}
\label{tab:kinfits}
\centering\small
\begin{tabular}{l l l l l l l l l l l }       
\hline\hline                 
GRB			& region 		& \multicolumn{2}{c}{Component 1 (narrow)}	& \multicolumn{2}{c}{Component 2 (narrow)}	& \multicolumn{2}{c}{Component 3 (broad)}& EW& F$_\mathrm{B}$/F$_\mathrm{N}$&V$_\mathrm{max}$	\\  
			&		& $\delta v$	& $\sigma$	& $\delta v$	& $\sigma$ & $\delta v$	& $\sigma$	& &\\ 
			&               & \scriptsize{km\,s$^{-1}$}&\scriptsize{km\,s$^{-1}$}&\scriptsize{km\,s$^{-1}$}&\scriptsize{km\,s$^{-1}$}&\scriptsize{km\,s$^{-1}$}&\scriptsize{km\,s$^{-1}$}& \scriptsize{\AA{}}& & \scriptsize{km\,s$^{-1}$}\\
\hline                       
GRB 020903  &main SF reg.   &3.6    &18.2 &40.2     &22.2     &--10.9    &58.1&--59$\pm$2 &0.46& 53\\
            &GRB reg.   &--40.2   &18.2 &--7.3     &18.2    & --18.3    &54.9&--61$\pm$3&0.36& 46\\
            &second SF reg. &--7.3    &18.2   &21.9     &24.1    &7.3    &51.6&--35$\pm$2&0.50&49\\
            &galaxy		&--7.3		&20.2	&25.6  &24.1		&--18.3		&54.9		&--62$\pm$3		&0.64&56\\
GRB 030329  &galaxy     &0    &21.6 &--     &--     & --74/+70    &unres./39.4 &--13$\pm$2&0.14&24\\
GRB 031203  &center     &0    &36.7  &--    &--     & --32.2   &105.8&--159$\pm$3&1.42&94\\
            &GRB reg.   &--1.9    & 35.3 &--    &--     & --28.5    &102.2 &--255$\pm$3&1.31&94\\	
             &broad reg. &7.0    & 36.3  &--     &--    & --26.8     &111.9&--182$\pm$2&1.79&99\\
            &narrow reg.&--13.3    &27.2 &--     &--     & --31.0    &103.9&--152$\pm$4&1.27&106\\
GRB 050826 &galaxy	&--44.1    &22.5  &22.9     &17.9     &  --19.4    &86.3   &--25$\pm$3&1.03&75\\
	&North/GRB  &--31.7    &18.0  &27.5     &5.0     &  --7.1    &75.7&--13$\pm$1&0.99& 62\\
            &center     &--28.2    &22.6   &28.2     &22.6      & --7.1   &75.7&--12$\pm$0.5&0.88&62\\
            &South      &--21.2    & 22.6 & 38.8    &22.6     & 0   &72.0 &--9$\pm$0.3&0.71&62\\
GRB 060218  &galaxy     &0    &11.3  &--24.8     &17.5   & --63.3  &40.9 &--27$\pm$5&0.30&13\\
GRB 100316D &Integ. 1   &--15.6    &5.6  &--     &--     & --11.2    &50.4& --25$\pm$1& 0.59&67  \\
            &Integ. 2   &--11.2    &10.0 &--     &--     &0    & 44.5 &--44$\pm$0.5& 0.23&64\\
            &Integ. 3/GRB &0    &16.8 &--     &--     &0   & 44.5&--184$\pm$2&0.20&60\\
            &Integ. 4   &44.9    &16.8 &--     &--    & 29.8   &39.6&--55$\pm$1&0.33&36\\
            &galaxy	&--3.0		&19.3	&--	&--	&12.0		&44.5		&--55$\pm$2	& 0.34&71\\
\hline                                
\end{tabular}
\tablefoot{The naming of the regions in the different hosts follows the one outlined in the line fitting plots, see Sect. \ref{sect:results}. We also list the H$\alpha$ EWs derived from the total line flux in each region and the flux ratio between narrow and broad component F$_\mathrm{B}$/F$_\mathrm{N}$. For GRBs\,020903 and 050826 we add the two narrow components for the total F$_\mathrm{narrow}$, for GRBs 030329 and 060218 we add the two components in the wings for the total F$_\mathrm{broad}$. V$_\mathrm{max}$ is defined in Sect. \ref{sect:discussionc} as abs($\Delta$v(narrow-broad) -- 0.5\,FWHM$_\mathrm{broad}$) \citep[see e.g.][]{Veilleux05, Arribas14}. For GRB\,020903 and 050826 we use $\Delta$v(narrow-broad) between the bluest narrow and the broad component, for GRB\,030329 we use the blue shifted additional component.}
\end{table*}

\subsection{GRB 020903}
 The host at $z=0.251$ is an irregular galaxy with either several SF regions or an interacting system as can be seen in HST images of the host \citep{Fruchter06, Svensson10} (see the contours in Fig. \ref{Fig:sample}). The galaxy is small with r$_{80}$ of 1.43\,kpc, has a stellar mass of log M*\,=\,8.87\,M$_\odot$ and a SFR of 1.7 M$_\odot$ yr$^{-1}$ \citep{Svensson10, Levesque10a}. \cite{Han10} found a tentative detection of Wolf-Rayet (WR) features in this galaxy. \cite{Levesque10a} also found a high ionization parameter for this galaxy indicative of a hard radiation field, which is not surprising given the presence of WR stars. 

We only detect H$\alpha$ in the FLAMES data, albeit with high S/N, and weak lines of [S{\sc\,ii}] in the integrated spectra. [N{\sc\,ii}] is not detected and we get a limit of 12+log(O/H)\,$<$\,8.1 for the N2 metallicity, in agreement with a metallicity of 7.98\,-\, 8.07 found by \cite{Levesque10b} and also the limit derived by \cite{Bersier06}. 

The galaxy is next to a system of two larger, interacting galaxies, however, as noted already by \cite{Soderberg04} they are at a redshift of $z=0.23$ compared to $z=0.25$ of the host galaxy ($\Delta v\,\sim\,$3900\,km\,s$^{-1}$) and hence not interacting with the host. At first sight the host of GRB 020903 seems to have a regular rotation curve with a $\Delta\,v$ of $\sim$\,60\,km\,s$^{-1}$.  The line width varies very little across the galaxy (see Fig. \ref{Fig:sample}), only the HII region east of the GRB region has a slightly higher value.

\begin{figure*}
\centering
	\includegraphics[width=15cm]{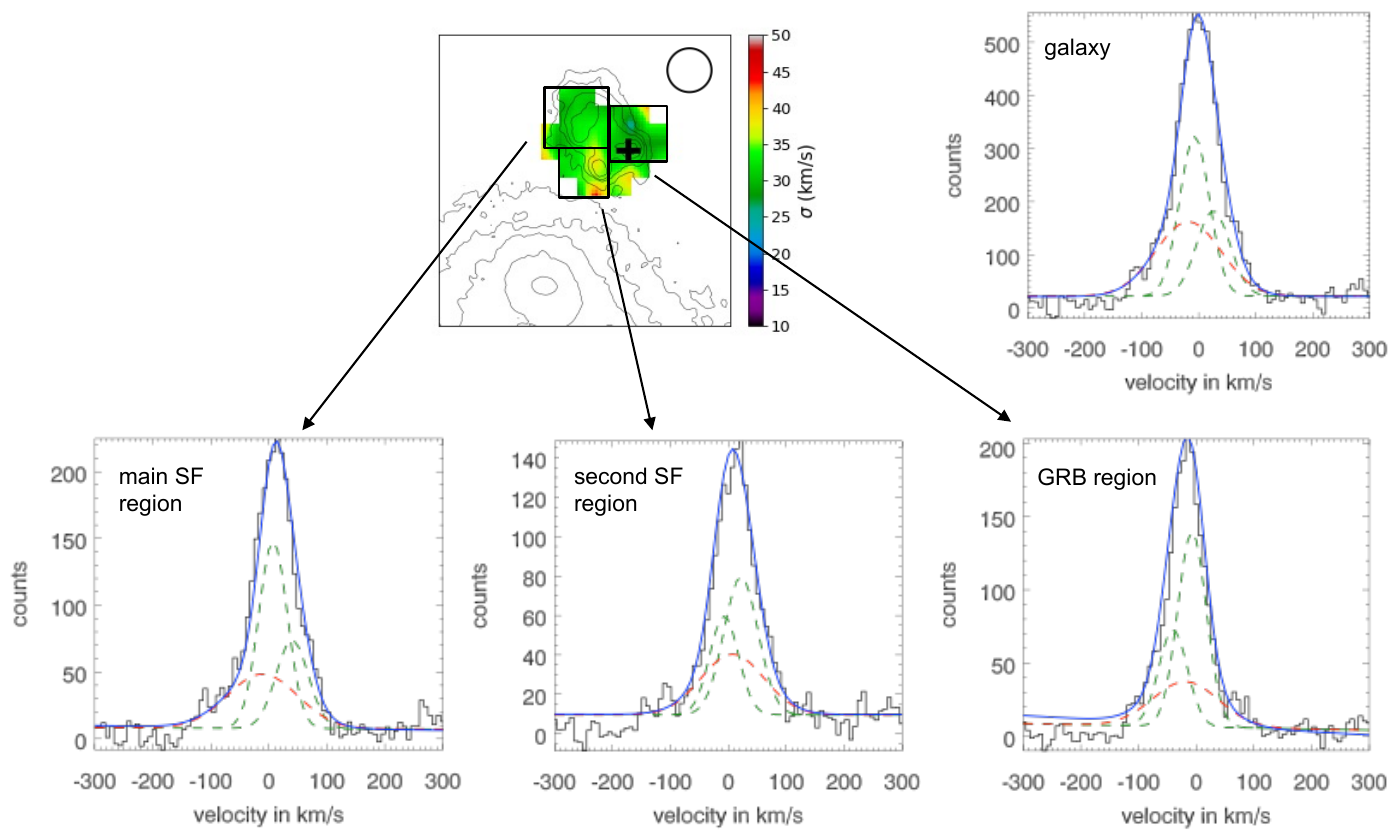}
    \caption{GRB 020903: Fits to H$\alpha$ for the GRB region, the main SF region, the SF region next to the GRB region and the entire galaxy. The circle in the $\sigma$ plot shows its nominal resolution.}
    \label{fig:020903_Hafits}
\end{figure*}

The emission line shows a slight asymmetry which we fit with a double narrow Gaussian profile. The ratio of the two profiles varies across the galaxy and seems to be associated with different parts of the galaxy (see Fig. \ref{fig:020903_Hafits}): The bluer component is related to the complex of SF regions in the north of the galaxy while the redder component is associated with the two SF regions in the South, one of which is the GRB region. The combination of these two components and their gradually varying relative strength across the galaxy give the appearance of an ordered rotation field, however, the velocity varies linearly across the FOV (see Fig. \ref{Fig:sample}), which is not expected for a rotating disk. Alternatively, this smoothly varying pattern could be indicative of an outflow along the WNW-ESE direction or an indication of a bar-like rotation.

In addition to the two narrow components, the line shows some excess emission in the blue wing which we fit with a weak broad component (see Fig. \ref{fig:020903_Hafits}). This component is most prominent in the part we call ``main SF region'' but less at the GRB site and the SF region next to the GRB site.

\subsection{GRB 030329}
The host is a compact, low mass (log M*$=$7.74 M$_\odot$, r$_\mathrm{80}=$\,1.03\,kpc, \citealt{Svensson10}) and low metallicity dwarf (12+log(O/H)\,$=$\,7.7\,--\,8.0,  \citealt{Thoene07, Levesque10a, Starling12}.) Even in high resolution HST images, the host seems to consist of a single SF region (see Fig. \ref{Fig:sample}). \cite{Ostlin08} determined a low age of 5\,Myr for the stellar population at the GRB site (assuming an instantaneous starburst) using high spatial resolution broad-band data from HST. \cite{Levesque10a} detect the Balmer series down to H8 together with [NeIII] emission, pointing to a young stellar population. The detection of [OIII] $\lambda$~4363\,\AA{}, which becomes very faint above 12+log(O/H)\,$>$\,8.0, confirms a very low metallicity for this galaxy.

We only detect H$\alpha$ both in individual spaxels and the integrated spectrum. [N{\sc\,ii}] is too weak at the metallicity of the host and the [S{\sc\,ii}] doublet is contaminated by bright sky emission lines. He{\sc\,i} 7065 would be in the observed spectral range but is not detected. This host, together with the host of GRB 060218, is one of only two GRB hosts where high resolution data of both absorption and emission lines are available.

\begin{figure}
	\includegraphics[width=9.2cm]{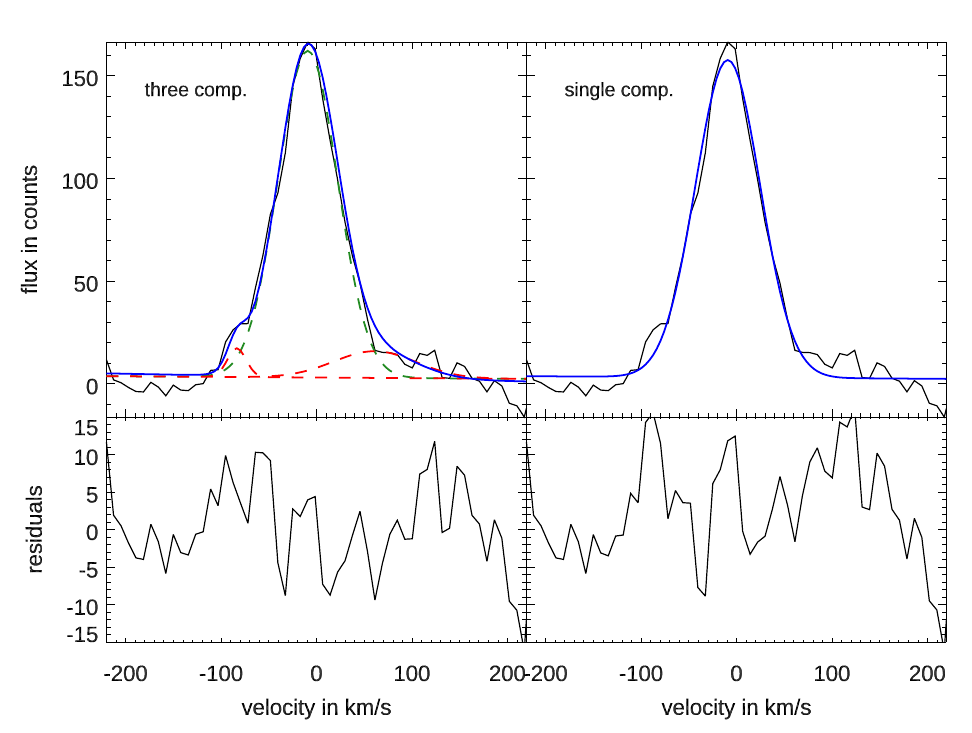}
    \caption{H$\alpha$ fits and residuals for the host of GRB 030329 using either a combination of a broad and a narrow component (left) or a single Gaussian (right). }
    \label{fig:030329_Hafits}
\end{figure}

The host of GRB 030329 has the smallest $\Delta v$ of all the sample (15\,km\,s$^{-1}$) and a uniform dispersion across the host. The S/N in the individual spaxels is too low to distinguish different components. In Fig.~\ref{fig:030329_Hafits} we fit the H$\alpha$ profile of the integrated galaxy spectrum using both a single Gaussian and two additional components in the blue and red wing of the line at velocities of --74 and +70\,km\,s$^{-1}$. There is a clear excess of emission in the blue wing, which seems to be present also in individual spaxels. The profile is slightly asymmetric with more emission in the blue part of the Gaussian, however, the data have too low S/N to constrain this further. 

Kinematics of the host had previously been analyzed in \cite{Thoene07} using longslit UVES/VLT spectra of the GRB afterglow. In those spectra, the emission lines show only a single Gaussian component while the absorption lines of Mg{\sc\,i} and M{\sc\,ii} span over 200 \,km\,s$^{-1}$ in velocity, blue-shifted compared to the emission line. This has been taken as a strong indication for a starburst wind in this galaxy. The possible excess emission in the wings of H$\alpha$ found in the FLAMES data would support this conclusion. Any faint component in the wings would have been missed in the UVES data as the afterglow continuum was still bright. 

\subsection{GRB 031203}\label{sect:031203_global}
The host is a compact, but luminous galaxy (log\,M*$=8.82M_\odot$), and has been observed at many wavelengths in multiple studies. \cite{Prochaska04} found a high extinction corrected SFR of 11 M$_\odot$yr$^{-1}$. The GRB was located near the galaxy center in the brightest region of the host. No individual SF regions have been identified and there are no HST images available. The galaxy hosts a young stellar population, indicated by the lack of a 4000\,\AA{} bump and the detection of high excitation MIR lines of Ne{\sc\,ii}+Ne{\sc\,iii} and [S{\sc\,iii}]+[S{\sc\,iv}] \citep{Watson11}. Tentative WR lines have been detected \citep{Han10}. Strong IR emission points to significant extinction but a high dust temperature ($\sim$ 70K) and the dust-to-stellar mass ratio is smaller than for other bright IR galaxies, suggesting dust properties different from local dwarfs \citep{Symeonidis}. \cite{Michalowski15} detected radio emission but offset toward the west of the galaxy and with a flat spectral slope suggesting a contribution from synchrotron self-absorption, indicative for a very young stellar population. The spectral slope rules out an AGN contribution, contrary to what was suggested by \cite{Levesque10a}.

\begin{figure*}[!ht]
\centering
	\includegraphics[width=15cm]{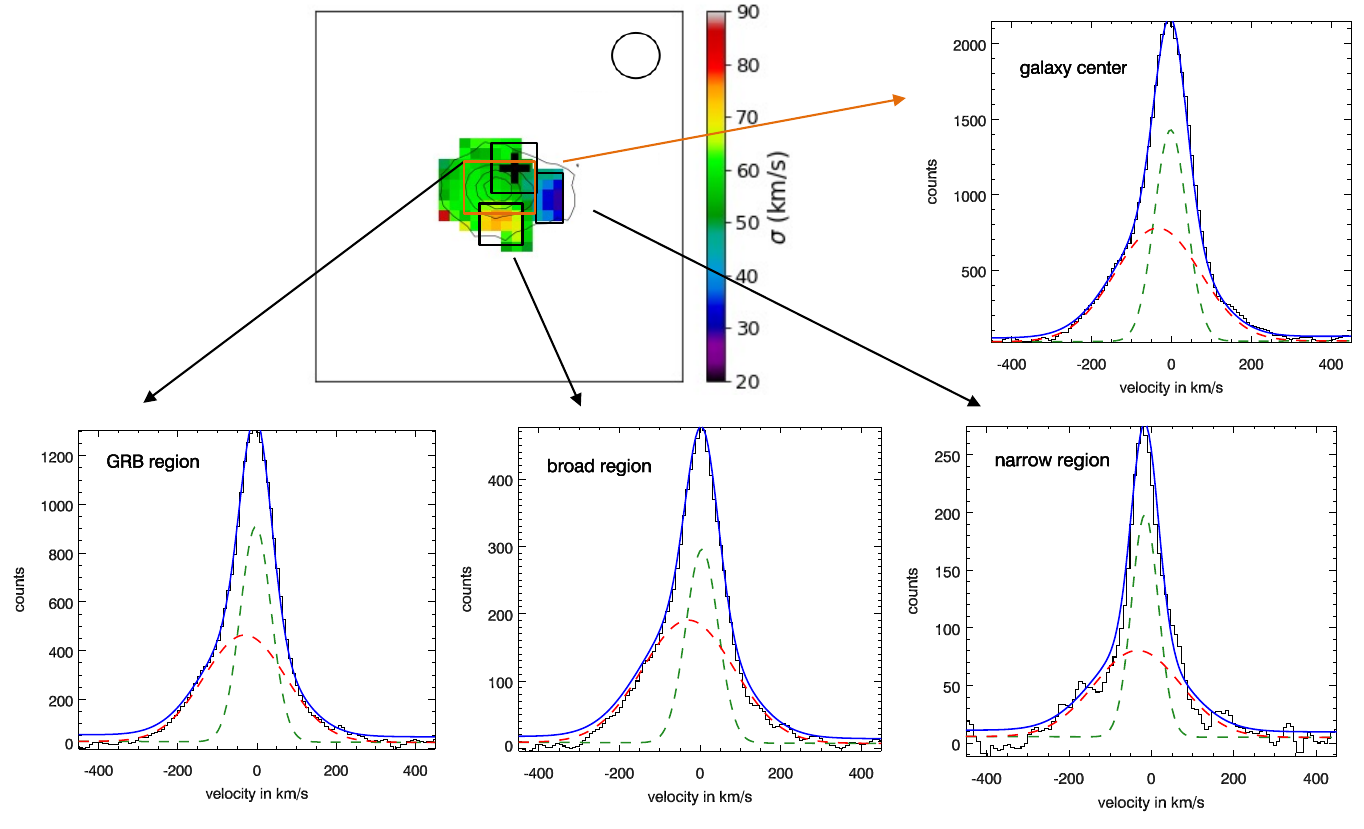}
    \caption{GRB 031203: Fits to H$\alpha$ for the GRB region, the region of high dispersion (``broad region'') and low dispersion velocity (``narrow region'') as seen in the dispersion map of the host. We add the profile of the integrated spectrum of the brightest 5$\times$6 spaxels in the galaxy center (orange rectangle) which we use for further comparison to X-shooter spectra of the host (see text). The circle in the $\sigma$ plot shows its nominal resolution.}
    \label{fig:031203_Hafits}
\end{figure*}

\begin{figure*}[!ht]
\begin{center}
	\includegraphics[width=15cm]{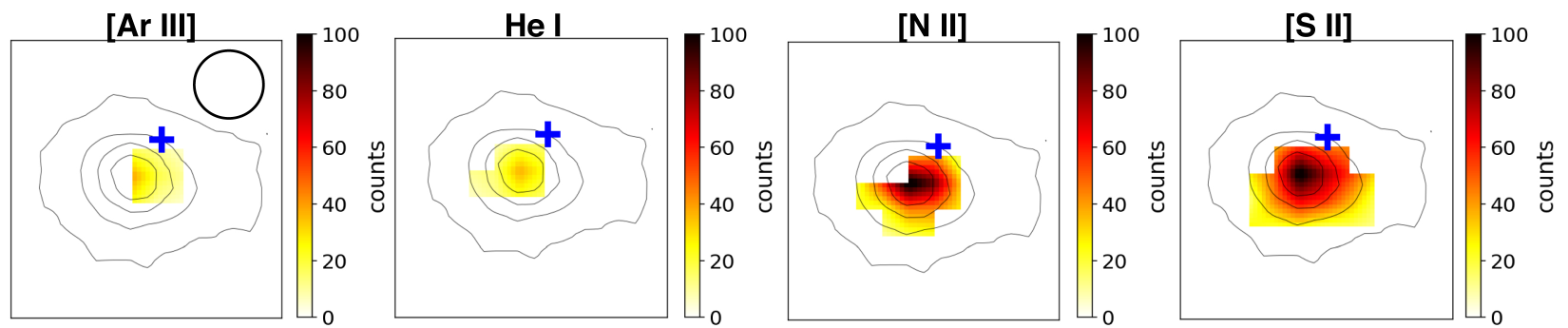}
	\includegraphics[width=15cm]{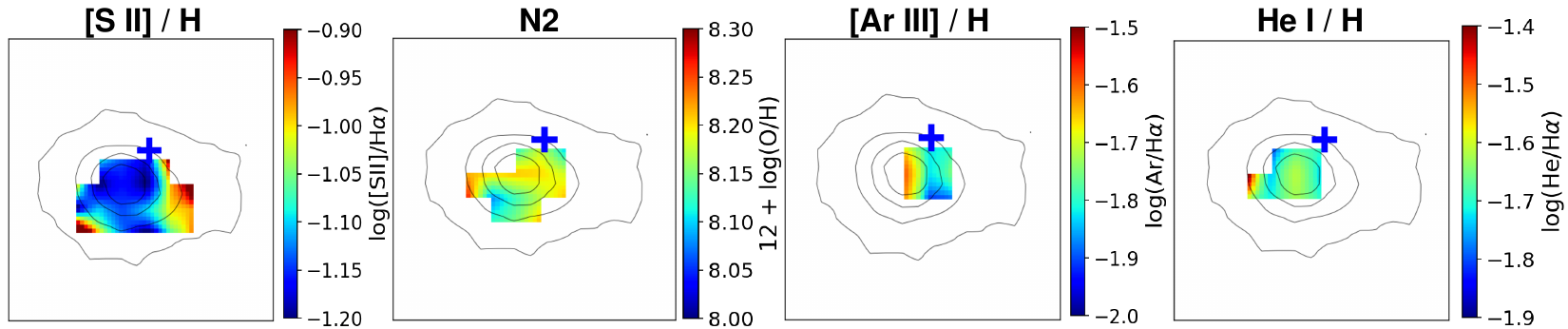}
    \caption{Top: Line maps of transitions detected in the spectra of GRB 031203, the position of the GRB is indicated by a cross. Bottom (left to right): [SII]/H$\alpha$, metallicity using the N2 parameter, [ArIII]/H$\alpha$ and HeI/H$\alpha$. The circle in the $\sigma$ plot shows its nominal resolution.}
    \label{fig:031203_abundances}
    \end{center}
\end{figure*}

We detect several emission lines even in individual spaxels: H$\alpha$, [N{\sc\,ii}], the [S{\sc\,ii}] doublet, [Ar{\sc\,iii}]~$\lambda$\,7136 and He{\sc\,i}~$\lambda$\,7065. The other two HeI transitions $\lambda$ 6678 and 7221 \AA{} are not detected, nor is the [O{\sc\,ii}]~$\lambda$\,7230, 7330 doublet. All of these lines were detected in X-shooter spectra of the host \citep{Guseva11, Watson11}. In Fig. \ref{fig:031203_abundances} we plot the line maps together with the ratio of [S{\sc\,ii}] /H$\alpha$, the metallicity as well as [Ar{\sc\,iii}]/H$\alpha$ and He{\sc\,i}/H$\alpha$. The lines all show a very similar distribution in flux with ratios varying by less than 0.3\,dex.

\begin{figure}[!ht]
\begin{center}
	\includegraphics[width=\columnwidth]{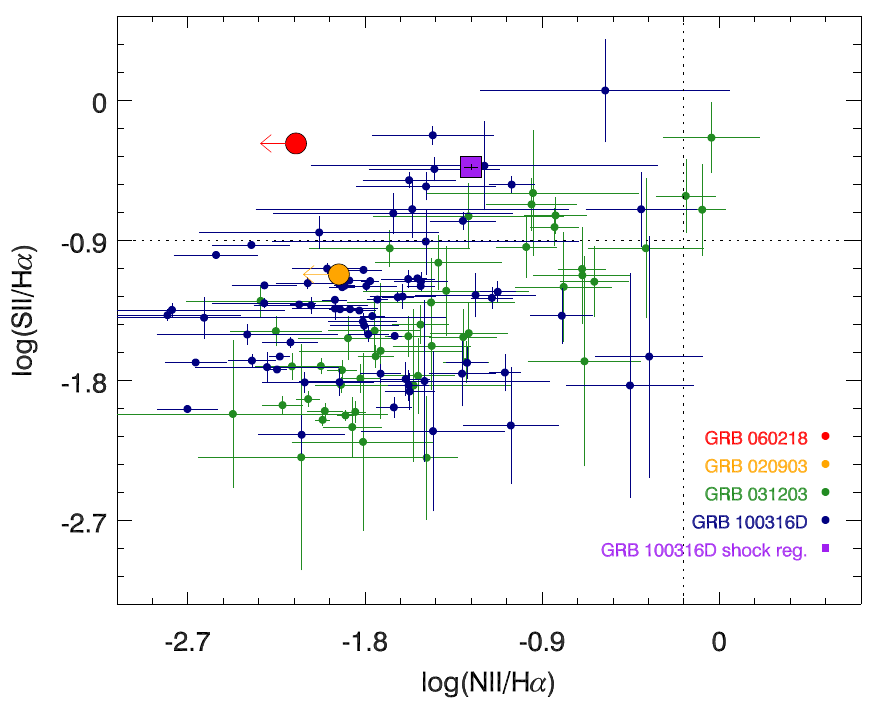}
    \caption{Ratios of [N{\sc\,ii}]/H$\alpha$ and [S{\sc\,ii}]/H$\alpha$ for individual spaxels in the hosts of GRB 031203 and GRB 100316D as well as the integrated spectra of GRB 020903 and GRB 060218. The latter two only have upper limits for  [N{\sc\,ii}]/H$\alpha$ since for [N{\sc\,ii}] only an upper limit can be measured. Dashed lines indicate the ratios below which ionization can be considered as the main source for the line excitation  \citep[see e.g.,][]{Westmoquette09b}.}
    \label{fig:ionplots}
    \end{center}
\end{figure}

The metallicity in different spaxels varies between 12+log(O/H)$\,=\,$7.9 and 8.4 with a median of 8.2 with a mean error of 0.08 (full range between 0.01 and 0.2). These values have been derived from the original data, not the interpolated map shown in Fig. \ref{fig:031203_abundances}. The GRB site and the south-east have somewhat lower metallicities while the center is marginally more metal-rich. [S{\sc\,ii}] /H$\alpha$ follows a pattern similar to [N{\sc\,ii}]/H$\alpha$ but with the lowest values concentrated toward the center of the galaxy. High [S{\sc\,ii}]/H$\alpha$ ratios can give some indications of shocked regions. Fig. \ref{fig:ionplots} shows the ratios of [N{\sc\,ii}]/H$\alpha$ and [S{\sc\,ii}]/H$\alpha$ for all spaxels where the four lines are detected with a S/N$>$3. Nearly all spaxels can be considered to be ionized by ionization from massive stars and not, for example, by shocks \citep{Westmoquette09b}.

The galaxy shows an ordered velocity field and a possible disk component. The total line-of-sight velocity difference across the galaxy, however, is only $\Delta\,v\,=$53\,km\,s$^{-1}$, which would point to a slowly rotating disk, a high inclination or a low mass, the latter of which is in conflict with the observations. The line width is rather uniform across the galaxy. However, there is a patch of low velocity width at the western end of the galaxy and a higher velocity region in the south, which we are going to investigate further in the following.

\begin{figure}[!ht]
\centering
\includegraphics[width=\columnwidth]{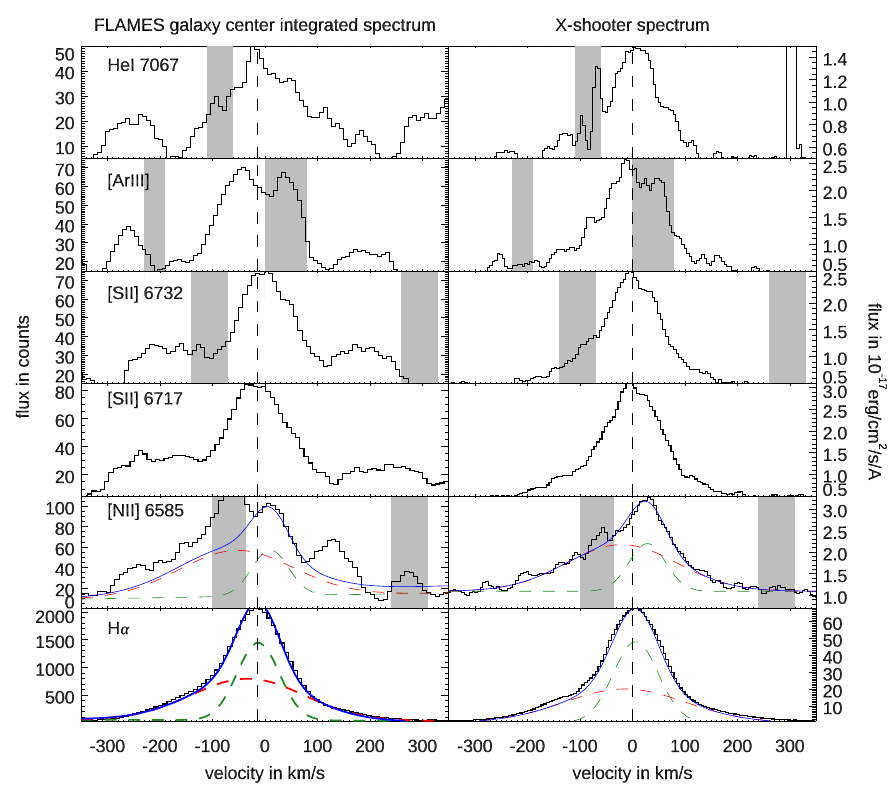}
    \caption{GRB 031203: Comparison between weak emission line profiles in a spectrum integrating over the central $6\times5$ spaxels of the galaxy center (left column) and the same lines in an X-shooter spectrum presented in \citet{Guseva11,Watson11}. The FLAMES spectra have been smoothed with a Gaussian kernel of 5 pixels in the spectral direction. Gray regions indicate contamination by residuals of atmospheric lines. For [N{\sc\,ii}]\,$\lambda$\,6585 we fit a combination between a narrow and broad component based on the X-shooter spectrum using the same parameters to fit the line in the FLAMES spectrum.}
    \label{fig:031203_Xsh}
\end{figure}

We extract integrated spectra of the GRB region, the high- and  low-$\sigma$ region and fit their line profiles (see Fig.~\ref{fig:031203_Hafits}). In all regions we clearly detect two components, a narrow, main peak with a $\sigma$ of $\sim$35 km\,s$^{-1}$ and a weaker, broad component with a $\sigma$ of $\sim$ 105 km\,s$^{-1}$, also detected in X-shooter (single slit) spectra of the host \citep{Guseva11,Watson11}. The relative strength of the component varies across the galaxy and explains the low- and high-$\sigma$ regions in the line width map, which is derived from a single Gaussian fit to H$\alpha$. The broad component is strongest in the high-$\sigma$ region in the South (``broad region'') and only slightly lower in the GRB region and weakest in the low-$\sigma$ region in the Western part of the galaxy (``narrow region''). Across all the galaxy, the broad component is blueshifted compared to the main emission component. In contrast to GRB\,030329, there are no afterglow spectra with absorption lines of this burst available to study a possible outflow in absorption.

\begin{figure}[!ht]
\centering
	\includegraphics[width=\columnwidth]{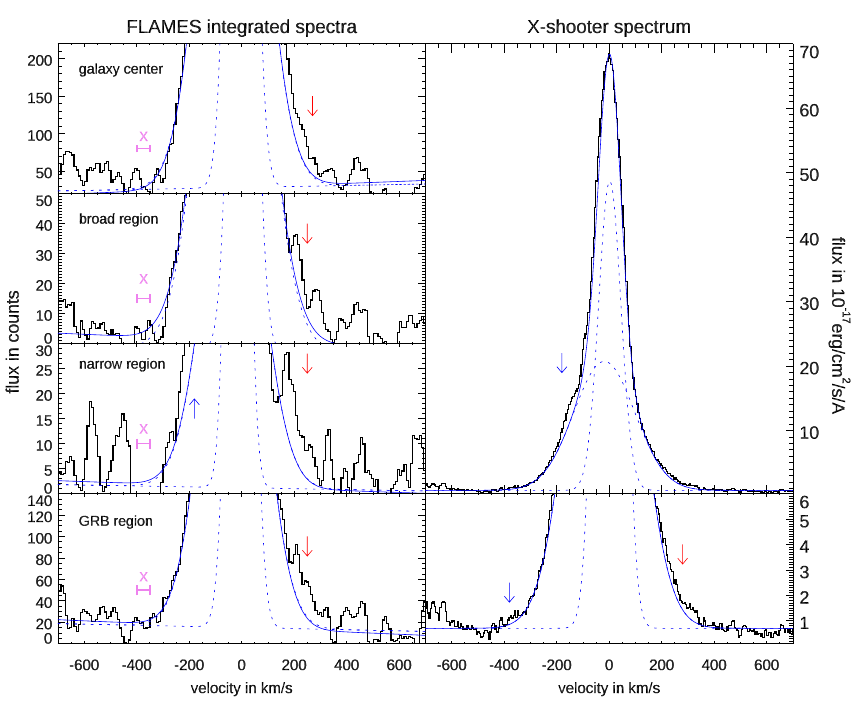}
    \caption{GRB 031203: Zoom into H$\alpha$ to look for the high velocity excess emission in both red and blue wings claimed to be detected in the X-shooter spectra by \cite{Guseva11}. {\it Left:} Integrated regions in the FLAMES spectra explained in Fig. \ref{fig:031203_Hafits}. {\it Right:} X-shooter spectrum. The scale for all panels (except the full size X-shooter panel) is $\sim$ 0.1 of the H$\alpha$ peak value. Excess emission components that we detect in both spectra are marked with arrows indicating with the same colors the same components detected in different spectra. The region where excess emission has been found in the blue wing of the X-shooter spectra but not in the FLAMES spectra is marked with a purple bar and cross. Zero velocity has been chosen as the centroid of the narrow emission component.}
    \label{fig:031203_highvcomp}
\end{figure}

We also look for possible broad components in the other detected lines of [Ar{\sc\,iii}], He{\sc\,i}, [S{\sc\,ii}] and [N{\sc\,ii}] (in the following named ``weak lines''). For this we made another integrated spectrum only comprising of the very central region of the galaxy ($6\times5$ spaxels) not to be dominated by noise. In Fig. \ref{fig:031203_Xsh} we plot the lines in velocity space compared to H$\alpha$. All weak lines have an irregular profile due to the low S/N and, except for [S{\sc\,ii}] $\lambda$6717, are affected by atmospheric lines. In the same figure we compare the lines to the ones from X-shooter data of the host presented in \cite{Guseva11,Watson11} (see Fig.~\ref{fig:031203_Xsh}). According to the finding chart in \cite{Guseva11}, the slit was oriented N-S with a width of 0.9--1.0 arcsec, covering a very similar region as the integrated spectra of the galaxy center. Despite the low S/N, all lines do show a broad wing that cannot be fit by a single Gaussian and excess emission on both sides and the emission stretches from $\sim$-150 to 150 km\,s$^{-1}$ and possibly even up to 300 km\,s$^{-1}$ in both red and blue. 

\cite{Guseva11} do not report a metallicity separately for the narrow and broad component, which they argue they are unable to provide due to the low S/N of the T$_e$ sensitive [O\,III]\,$\lambda$\,4363 line. Why they do not try to fit multiple components to [N{\sc\,ii}]\,$\lambda$\,6585 is not mentioned. Although the [N{\sc\,ii}] line is affected by an atmospheric line, we fit a double component similar to H$\alpha$ to the X-shooter spectra and use the same parameters to constrain the line in the FLAMES spectra of the galaxy center, which is more affected by the atmospheric line in the blue wing (see  Fig.~\ref{fig:031203_Xsh}). Despite the atmospheric contamination, the broad component relative to the narrow component is stronger in [N{\sc\,ii}] than in H$\alpha$. From the fits to the components in the X-shooter spectra we derive N2 metallicities of 12+log(O/H)$\,=\,$8.03$\pm$0.20 for the narrow component and 8.19$\pm$0.16 for the broad component. Although the metallicities are consistent within errors, there might be a small trend for a higher metallicity of the broad component.



\cite{Guseva11} claim excess emission beyond the broad-narrow profile at $\sim$ --400 km\,s$^{-1}$ and +350 km\,s$^{-1}$ in H$\alpha$, H$\beta$ and [O{\sc\,iii}]. In Fig. \ref{fig:031203_highvcomp} we plot a zoom of H$\alpha$ in the three regions studied in Fig. \ref{fig:031203_Hafits} and the extraction of the central 5$\times$4 spaxels (see last paragraph). We are not able to recover the excess emission at --400 km\,s$^{-1}$ but we do see emission redward of the broad emission component at 200 -- 300 km\,s$^{-1}$ in all regions, coincident with the excess emission in the X-shooter spectra (right panel). In some spectra this emission even looks like a narrow extra component in the red wing. Surprisingly, this component has the highest relative strength compared to the rest of the line in the ``narrow region'' with the lowest contribution of the broad component. In this region, we also clearly see an extra ``shoulder'' in the blue wing of H$\alpha$ at --150 to --200 km\,s$^{-1}$, also apparent in the X-shooter spectra plotted in Fig. \ref{fig:031203_highvcomp} (right top panel in the figure) but not in the fit shown in \citet{Guseva11}, probably due to smoothing of the spectra.  The reason for not detecting it in other regions might be that this extra component is blended with the broad component due to its higher line width in those regions. 

\subsection{GRB 050826}
The host of GRB 050826 ($z=0.297$) is visually a compact galaxy but has the highest stellar mass of the sample with log\,M*$=9.79M_\odot$. There is some discrepancy in the literature on the SFR inferred from different indicators: While \cite{Svensson10} find a moderate SFR of 1.39\,M$_\odot$/yr based on U-band photometry, \cite{Levesque10b} derive a somewhat higher value of 2.9\,M$_\odot$/yr from the H$\alpha$ emission line. \cite{Levesque10b} determined a supersolar metallicity for the host galaxy based on the R$_{23}$ parameter, taking the upper branch of the solution. \cite{Mirabal050826} detected also [O{\sc\,ii}] from the host and report the detection of the afterglow. No SN component has been detected due to the late confirmation of the afterglow. The initially reported transient (which later turned out to be the afterglow) was not coincident with the X-ray afterglow position \citep{HalpernGCN1}, and not before Feb. 12, 2006 it was finally confirmed as the optical afterglow \citep{HalpernGCN2}.

We only detect H$\alpha$ in the integrated spectrum of the galaxy. As this is the highest redshift host of the sample, despite its high mass and luminosity, the S/N of even H$\alpha$ is low. H$\alpha$ is also next to a bright atmospheric emission line in the very blue wing of the line, hence most of the profile is not affected. Given the stated high metallicity of the host by \cite{Levesque10b}, we should easily detect [N{\sc\,ii}]. However, our reanalysis resulted in a revised metallicity for this host (see Appendix B).  

\begin{figure*}
\centering
	\includegraphics[width=15cm]{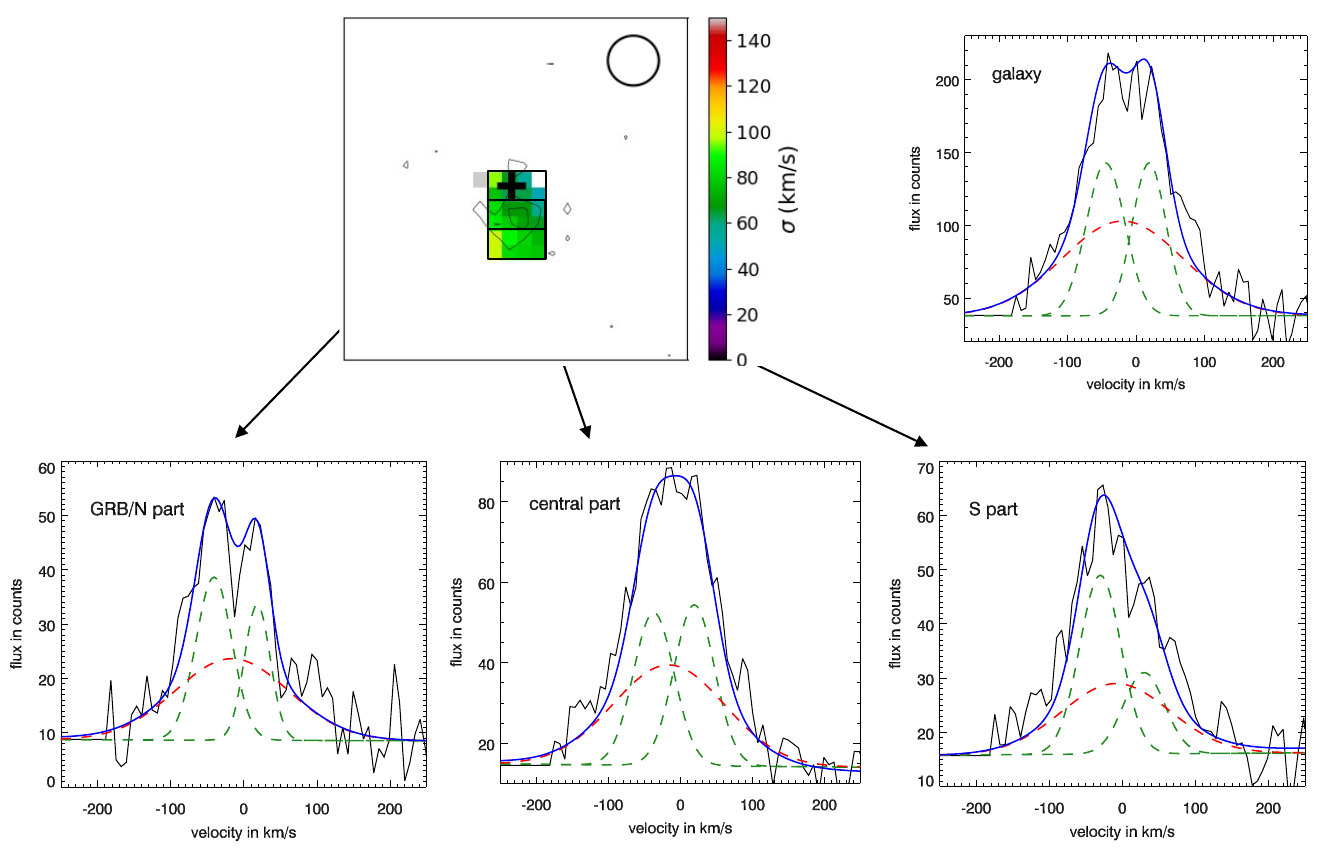}
    \caption{GRB 050826: Fits to H$\alpha$ for three different regions in the host using a triple Gaussian component. We also show the profile and fit to the integrated spectrum of the host. The circle in the $\sigma$ plot shows its nominal resolution.}
    \label{fig:050826_Hafits}
\end{figure*}

H$\alpha$ shows a double peak in part of the host (see the discussion below and Fig.~\ref{fig:050826_Hafits}). To be able to derive a velocity map fitting a single Gaussian, we first smooth the cube in wavelength with a Gaussian kernel of 7 pixels before fitting the resulting single peaked line. The velocity field of the galaxy is very non-uniform with no clear rotation and the difference in velocity across the galaxy is only 43\,km\,s$^{-1}$. The width of the line has a high $\sigma$ of 73\,km\,s$^{-1}$ owing to the asymmetric and double line profile. The line width also shows some trend from lower velocities around 50 km\,s$^{-1}$ on the N-W side, where the GRB is located, up to almost 100 km\,s$^{-1}$ on the opposite side of the galaxy. 

In Fig. \ref{fig:050826_Hafits} we fit H$\alpha$ to three different regions in the host and in the integrated spectrum of the entire galaxy. The H$\alpha$ line is broad at the center of the galaxy, asymmetric in the South dominated by  the bluer component and clearly double-peaked in the northern part at the position of the GRB, which is also apparent in individual spaxels (hence it is not an artifact). This behavior explains the strange-looking velocity field, which is due to the shifting distribution of the double component across the host, similar to the host of GRB 020903.

We first fit the line with a double Gaussian component with a velocity difference of $\delta$v$\,=\,$60 km\,s$^{-1}$ and $\sigma$ of 6--17 km\,s$^{-1}$ for both components. There is clear excess emission in both wings, hence we include a third, broad, component with a $\sigma$ of 75 km\,s$^{-1}$. A fit to this component with a single Gaussian might not be optimal, alternatively, two narrower components could be fitted in the blue and red wings as we did for the host of GRB\,030329. The centroid of the different components basically does not change, only their relative contributions. The broad component is present everywhere at a similar relative strength compared to the double narrow component. In the combined spectrum of the host, the broad component is less evident and the spectra could be equally well fit with a double component with a slightly higher $\sigma$ than the narrow components in the triple component fit, underlining the importance of spatially resolved spectroscopy (see also Sect. \ref{sect:resolved}).

\subsection{GRB 060218}

\begin{figure}[h]
	\includegraphics[width=8.95cm]{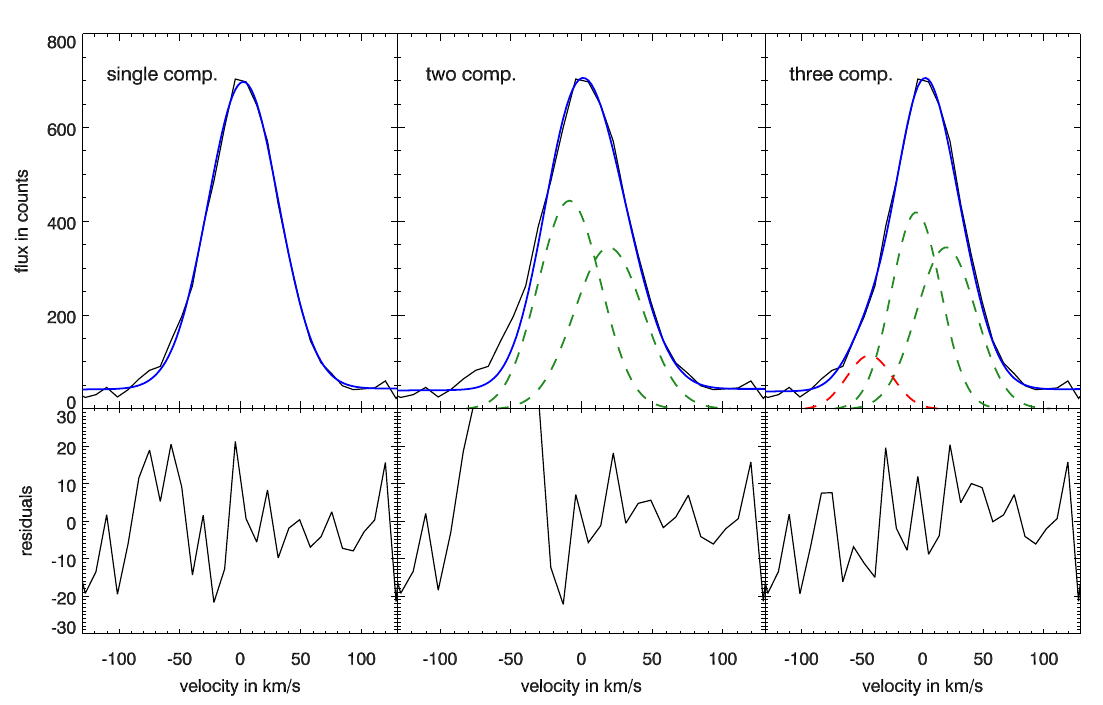}
    \caption{H$\alpha$ fits and residuals for the host of GRB 060218. Left: Single Gaussian component, middle: two Gaussians with the same distance as the two components found in the afterglow spectra of GRB\,060218 in \cite{Wiersema07}. Right: Additional component to fit the higher velocity emission in the blue wing of the emission line.}
    \label{fig:060218_Hafits}
\end{figure}

The host is the least luminous (M$_B$\,=\,--15.9\,mag, log~M*\,=\,7.4\,M$_\odot$), smallest (r$_\mathrm{80}=$0.55\,kpc) and most metal poor host detected (12+log(O/H)\,$=$\,7.6 or $\sim$\,0.1 Z$\odot$, \citealt{Wiersema07,Kewley07}). Even HST imaging \citep{Starling12} does not resolve different HII regions. We detect only H$\alpha$ in the individual spaxels and marginally detect [S{\sc\,ii}]  in the integrated spectrum of the host. The limit on the ratio of log( [N{\sc\,ii}]/H$\alpha$)\,$=$\,--1.87 from the integrated spectrum implies a metallicity limit of 12+log(O/H)\,$<$\,7.88, consistent with the value determined in other longslit data. 

The galaxy has no regular velocity field and does not show any sign of rotation. The dispersion is nearly constant, only at the S-W edge there might be a region with slightly higher width. The GRB lies in one of the regions with the lowest $\sigma$, just outside the brightest region of the galaxy. Since the galaxy is very compact and low-mass, the absence of a regular velocity field is not surprising. 

\cite{Wiersema07} obtained high-resolution UVES data at maximum light of the SN and find two kinematical components in NaD absorption separated by 24\,km\,s$^{-1}$ and a corresponding double component in [O{\sc\,iii}] emission (H$\alpha$ was out of the range). We tried to recover these two components in the H$\alpha$ line of the global spectra of the galaxy (see Fig.~\ref{fig:030329_Hafits}) and indeed the emission line does show a very small asymmetry. Fitting a single Gaussian does give a reasonable fit with only a small excess in the blue wing of H$\alpha$. We then fit a double component with a velocity difference of $\delta$v$=$24.8\,km\,s$^{-1}$ and $\sigma$ of 12 and 15 km\,s$^{-1}$ respectively. These values are similar to the ones found in \cite{Wiersema07}, who get a $\delta$v of 21.6 km\,s$^{-1}$ and $\sigma$ of 15 and 20 km\,s$^{-1}$, respectively. This double component initially gives a worse fit, but adding a small component in the blue wing with a $\sigma$ of 8 km\,s$^{-1}$ results in the best residuals.

The presence of two main emission components would probably have been missed in our FLAMES spectra without the information from the UVES spectra. The absence of the third component in \cite{Wiersema07} might have either been due to the higher continuum emission from the GRB-SN or its association with a part of the host not covered by the UVES spectra. While a single Gaussian profile has a similarly good fit as the triple profile, once fitting the double profile as in \cite{Wiersema07}, a third component is needed to account for the additional emission in the blue wing. Since our S/N is too low, we cannot extract spectra from different parts of the host to see whether this is associated with a specific region.

\subsection{GRB 100316D}

\begin{figure*}[h!]
\begin{center}
\includegraphics[width=15cm]{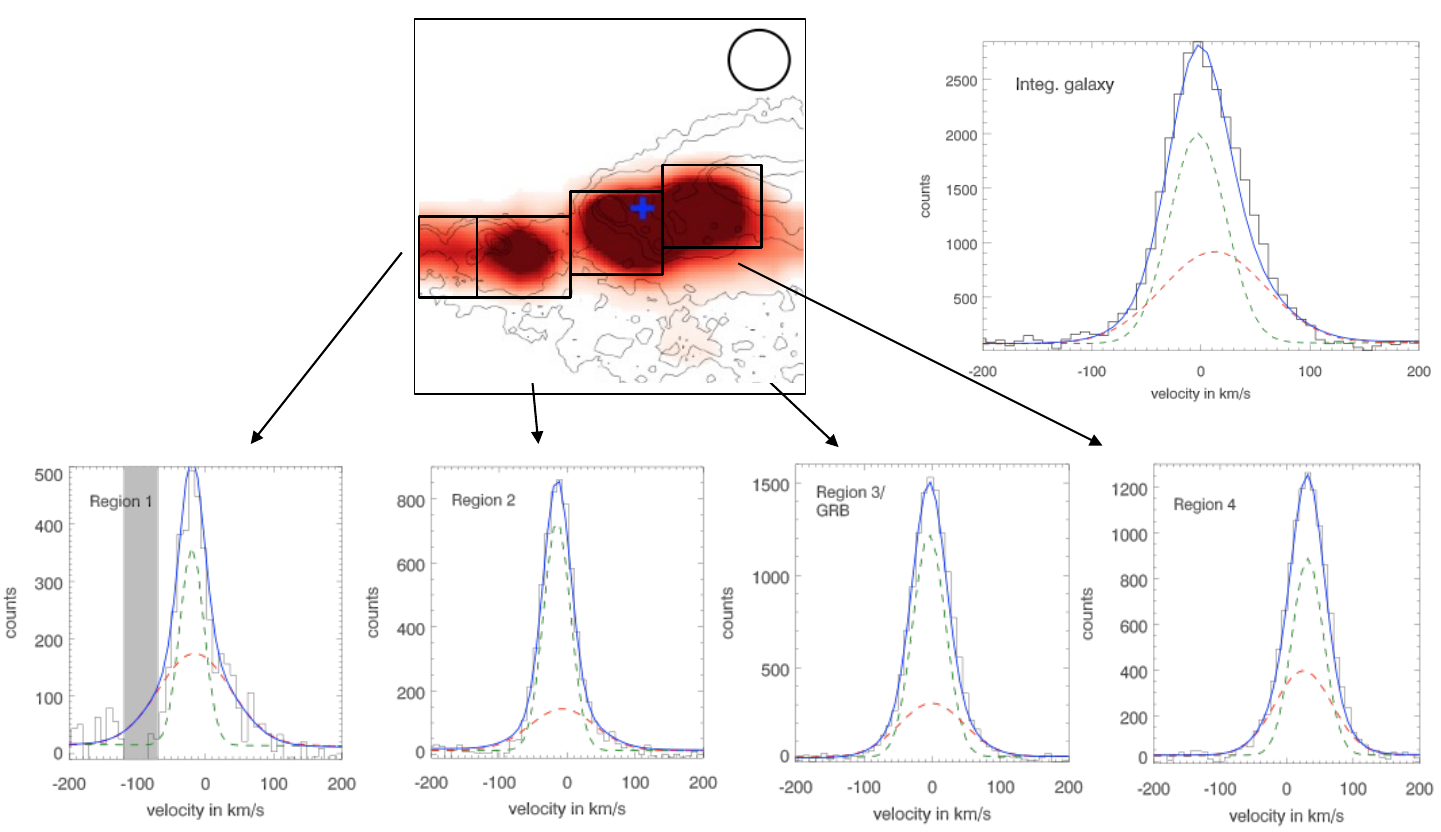}
\caption{GRB 100316D: Fits to H$\alpha$ in four integrated HII regions in the host and the integrated spectrum of all the spaxels in the FOV (which cover about 1/3 of the galaxy). The circle in the $\sigma$ plot shows its nominal resolution.}
\label{fig:100316D_Hafits}
\end{center}
\end{figure*}

The host galaxy is a low-mass (log~M\,$=$\,8.93 M$_\odot$), irregular, highly star-forming galaxy \citep{Starling12, Levesque11} with several bright SF regions and its close distance allows for a detailed, resolved analysis (the angular size of the host is $\sim$\,12 arcsec). The metallicity is low with 12+log(O/H)\,$=$\,8.0\,--\,8.2 \citep{Levesque11, Izzo17a} and it has total SFR of 1.2~M$_\odot$/yr (\citealt{Izzo17a}, henceforth I17). The GRB is located at the edge of the brightest and most extreme star-forming region in terms of SFR and metallicity. 

The FOV of FLAMES only covers about half of the galaxy due to an error in the observational setup, but does include most of the bright SF regions in the northern part together with the GRB region. In individual spaxels we detect H$\alpha$, [N{\sc\,ii}] and the [S{\sc\,ii}] doublet at high S/N. I17 used the FLAMES data presented here for the kinematic analysis of the host. The MUSE data reveal a large number of emission lines including some not previously detected in GRB hosts such as [N{\sc\,i}] and  [Fe{\sc\,ii}]. The detection of HeI emission implies a very young stellar population and I17 derive an age of 5\,Myr for the population at the GRB site.

\begin{figure*}[h!]
\begin{center}
\includegraphics[width=15cm]{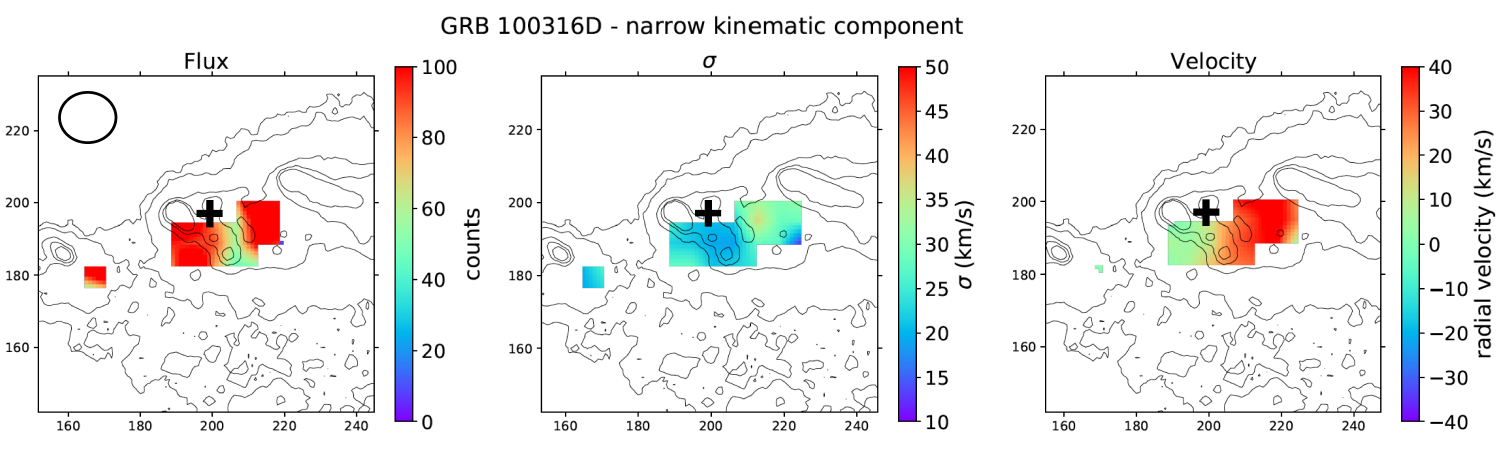}\\
\includegraphics[width=15cm]{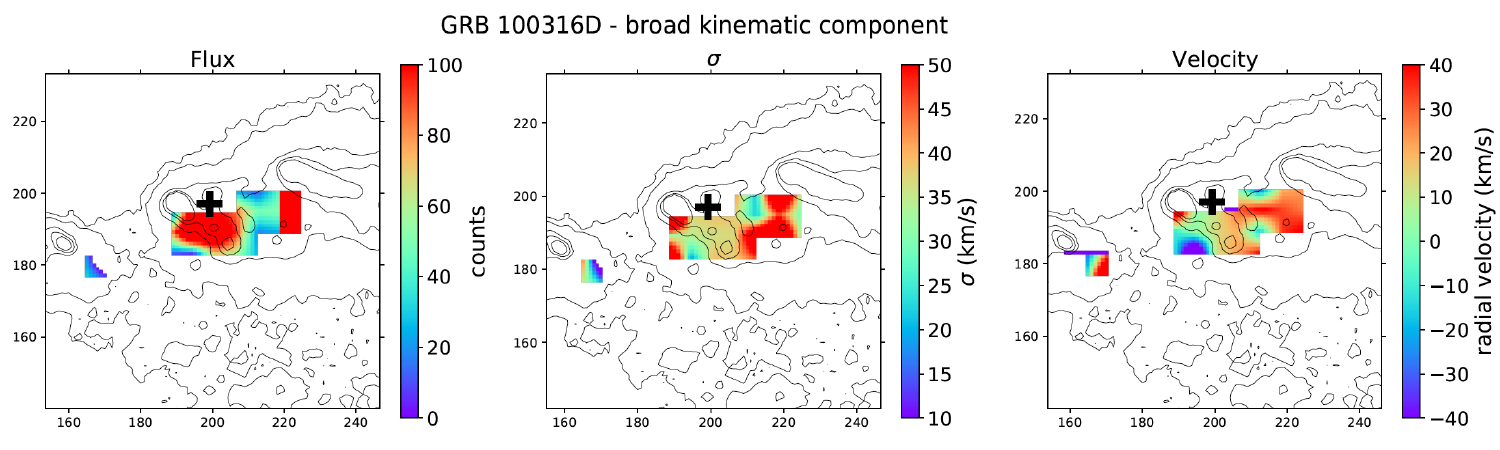}
\caption{GRB 100316D: Flux (in arbitrary units), velocity width and velocity field of the narrow (top) and broad (bottom) emission line components. The circle in the $\sigma$ plot shows its nominal resolution.}
\label{fig:100316D_compflux}
\end{center}
\end{figure*}

\begin{figure}[h!]
\begin{center}
	\includegraphics[width=\columnwidth]{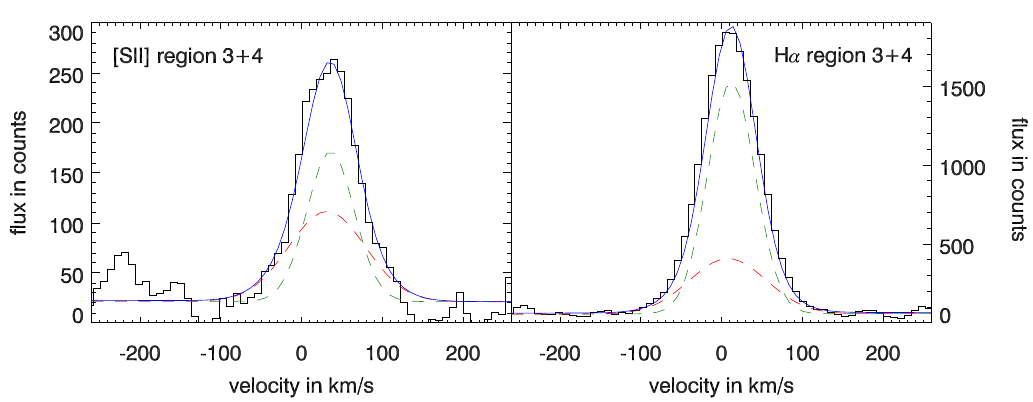}
    \caption{GRB 100316D: Fit of a double component to [S{\sc\,ii}]$\lambda$ 6717 in a region combining the SF region 3 (next to the GRB site) and SF region 4 and the corresponding fit to H$\alpha$ in the same regions. }
    \label{fig:100316D_SIIfit}
    \end{center}
\end{figure}

\begin{figure*}[h!]
\begin{center}
	\includegraphics[width=18cm]{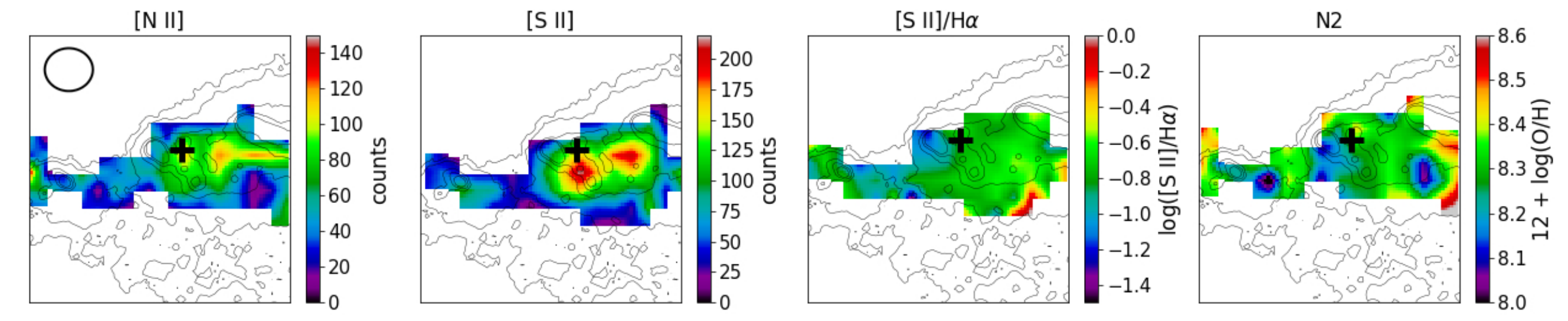}
    \caption{Line maps of [\ion{N}{II}] and the [\ion{S}{II}] doublet in the spectra of GRB 100316D, [S{\sc\,ii}]/H$\alpha$ and metallicity using the N2 parameter \citep{MarinoZ}. The circle in the $\sigma$ plot shows its nominal resolution.}
    \label{fig:100316D_abundances}
    \end{center}
\end{figure*}

The host of GRB 100316D is the only galaxy with sufficient S/N to make 2D maps of different emission components. H$\alpha$ shows a clear broad emission component in most of the spaxels. For spaxels with an H$\alpha$ line flux of $>$ 80 counts we fit a double component using the PAN line fitting tool (see Fig. \ref{fig:100316D_compflux})  \footnote{Developed by Rob Dimeo, \url{http://ifs.wikidot.com/pan}}. The broad component is strongest in the bright SF region next to the GRB region. The narrow component shows a regular velocity field of a rotating disk while the broad component is more chaotic. The velocity width of the narrow component is lowest at the center of the SF regions and higher on the edges while the width of the broad component is more erratic, which might be a simple effect of S/N. 

In Fig. \ref{fig:100316D_Hafits} we further study the integrated  H$\alpha$ profile of four SF regions, but do not find any additional components. In individual regions, the broad component is blueshifted compared to the narrow component, however, in the integrated host spectrum, the profile is skewed to the red and the broad component appears redshifted compared to the main emission peak. This is an artifact from the combination of different regions with different relative contributions and the velocity shift due to rotation of the galaxy.

I17 concluded from MUSE data that the GRB site is metal poor and next to the most extreme region in the host. The spatial resolution of the FLAMES data is slightly higher due to better seeing conditions ($\sim$0.8'' for the FLAMES data vs. $\sim$1.1'' for MUSE). The N2 metallicity follows the same pattern as in I17 with the lowest metallicity (12+log(O/H)$\,=\,$8.1) in the bright SF region next to the GRB site. In the FLAMES data we also see a metal poor region with the same metallicity to the South-West of the GRB region. I17 only show a combined [S{\sc\,ii}+N{\sc\,ii}]/H$\alpha$ map, which is rather uniform in the part covered by the FLAMES data and higher values in a region they consider to be affected by shocks. Our [S{\sc\,ii}]/H$\alpha$ map shows no large variations but has particularly low values in the bright SF region next to the GRB site. 

Also here we search for indications of shock excitation in the [S{\sc\,ii}]/H$\alpha$ vs. [N{\sc\,ii}]/H$\alpha$ plot (see Fig.\ref{fig:ionplots}). Most spaxels are not affected by shock excitation. I17 found a region possibly affected by shocks in the Western part of the galaxy, outside the FOV of the FLAMES datacube. For comparison we plot the values found from an integrated spectrum of this possibly shocked region and the values are indeed indicative of shocks, however only in [S{\sc\,ii}]/H$\alpha$, while the value for [N{\sc\,ii}]/H$\alpha$ is in a region that could be excited only by ionizing radiation.

We furthermore attempted to determine the metallicity of the broad and narrow component separately. Due to the low S/N of [N{\sc\,ii}]\,$\lambda$\,6585 we use  [S{\sc\,ii}]\,$\lambda$\,6717. As we see in Fig. \ref{fig:100316D_SIIfit} the broad component is stronger compared to the narrow component than in H$\alpha$. The fluxes of both components are very similar for [S{\sc\,ii}] while the narrow component of H$\alpha$ has a 2.5 times higher flux than the broad component. Deriving a concrete metallicity value is hindered by two problems: The S2 parameter requires both lines of the [S{\sc\,ii}] doublet since the ratio depends on the electron density which varies between $\sim$\,1.5 and 0.5 for low and high densities, respectively. However, the $\lambda$6732 line does not have enough S/N to fit both components. Second, the S2 metallicity has a weaker correlation with the N2 parameter. As we measure a ratio of $\sim$\,1.5 for both the peak and the total flux ratio, we assume a low electron density and would therefore obtain metallicities of 8.3$\pm$0.1 and 8.7$\pm$0.2 for the narrow and broad component respectively, using the metallicity based on the S2 index from \cite{Yin07}.


\section{Discussion}
\subsection{GRB host classification and velocity fields}\label{sect:discussion}
Our sample of GRB hosts spans a wide range of stellar masses (log M*$=$7.5 and 9.9 M$_\odot$) and sizes (r$_\mathrm{80}$ = 0.6 to 12\,kpc). In the following, we try to classify the galaxies using not only the mass, size and luminosity but also their velocity field. 

BCDs show all kinds of velocity fields from ordered, disk-like rotation to highly disturbed velocity fields \citep[see e.g.][]{Ostlin01, BlascoHerrera12, Cairos15, Cairos17}. Very compact dwarfs such as GPs \citep{Lofthouse17} often do not show a regular rotation field. dIrrs are considered to be on the low-mass end of disk galaxies and have a regular velocity field with the only difference being the absence of a bulge component \citep[e.g.][]{Swaters09}. 

A commonly used criterion to distinguish between dispersion vs. rotation dominated galaxies is the following: v$_\mathrm{shear}$/$\sigma_0$ $>$1 with v$_\mathrm{shear}=$\,0.5\,$\Delta\,v$, where $\Delta\,v$ is the difference between the minimum and maximum velocity of H$\alpha$, and $\sigma_0$ the flux weighted average of the line width in all spaxels with S/N H$\alpha >$3 (see values in Tab. \ref{tab:globalkin}). Using $\sigma_0$ and v$_\mathrm{shear}$ we can derive a dynamical mass for the hosts. For spherical, relaxed systems the dynamical mass derived from the virial theorem is
\begin{equation*}
M_\mathrm{dyn,\sigma}={4 r_\mathrm{50} \sigma^2\over{G}}
\end{equation*}
For rotation dominated systems the mass is derived as
\begin{equation*}
M_\mathrm{dyn,rot}={2 v_\mathrm{rot}^2 r_\mathrm{50}\over{G}}
\end{equation*}
\citep{Bellocchi13} where v$_\mathrm{rot}$=0.5$\Delta\,v$, identical to v$_\mathrm{shear}$ derived above. M$_\mathrm{dyn,rot}$ technically comprises the mass within the half light diameter using the velocity at the half-light or effective radius while we use half the maximum velocity spread for v$_\mathrm{rot}$. At the resolution of our dataset we can safely set  v$_\mathrm{rot}$ $\sim$ v$_\mathrm{eff}$. In Tab. \ref{tab:globalkin} we list the values from both methods, for any further calculations we use  M$_\mathrm{dyn,\sigma}$ for the hosts of GRBs 030329, 060218, 020903 and 050826 and M$_\mathrm{dyn,rot}$ for the hosts of GRBs 100316D and 031203. We do not apply any correction for inclination, which would affect the value of M$_\mathrm{dyn,rot}$.

The two smallest hosts, those of GRB 030329 and GRB 060218, clearly fall in the category of BCDs. Both galaxies show almost no rotation and are dispersion dominated. The stellar masses are much lower than those from M$_\mathrm{dyn,\sigma}$, which could point to a more turbulent system. The contribution of the broad component is small and likely does not play a large role in determining $\sigma_0$ (see F$_\mathrm{broad}$/F$_\mathrm{narrow}$ in Tab. \ref{tab:kinfits}). 

The hosts of GRB 020903 and GRB 100316D are dIrr with many different SF regions. The host of GRB 100316D has r$_{80}$=4\,kpc and shows intense SF around the GRB site but a low surface brightness in the rest of the galaxy. It also has the most pronounced disk-like rotation of our sample. I17 found that the galaxy is similar to other dIrr, being dominated by the disk rotation at small radii but by dark matter (DM) at larger radii. They concluded that the galaxy might not be completely virialized and suggested a close encounter with another (unknown) neighbor. This could explain the ongoing starburst in part of the galaxy ($\sim$40 \% of the SF is located in the 25\% of the galaxy covered by FLAMES). The host of GRB 020903 is compact, has no regular rotation curve, and is dominated by two main emission components close in velocity space that shift in strength across the galaxy. The host has v$_\mathrm{shear}$/$\sigma_0$ $\sim$1, note, however, that it is difficult to determine v$_\mathrm{shear}$ due to the double emission component. 

Finally, the hosts of GRB 031203 and GRB 050826 are on the high mass end of dwarf galaxies. \cite{Watson11} classified the host of GRB\,031203 as a BCD due to its hard radiation field, high SF rate and absence of large amounts of dust. The host does show a clear rotation curve but a very low v$_{\max}$ of only $\sim$25\,km\,s$^{-1}$. M$_\mathrm{dyn,\sigma}$ and M$_\mathrm{dyn,rot}$ match very well but are lower than the stellar mass. This could be due to a high inclination of the disk, however, the wind component is strong, which would speak for a relatively face-on system. With the lack of high resolution imaging it is difficult to determine its exact morphology. Formally, the host of GRB 050826 is dispersion dominated, but the two narrow components complicate the applicability of this criterion. The double component and absence of a clear rotation field could be an indication for a merger. Also M$_\mathrm{dyn,\sigma}$ is much higher than M$_\mathrm{dyn,rot}$ due to the ``artificially'' large $\sigma$ from the double component. Again, the lack of high-resolution imaging prevents further conclusions.

\subsection{GRB location}
The location of the GRB is usually not at the region with the highest H$\alpha$ flux, but often close to it. \cite{Fruchter06} and subsequent studies \citep[e.g.][]{Kelly08, Svensson10, Lyman17} concluded from high spatial resolution HST imaging that GRB-SNe have a higher correlation with the brighter and bluer regions than any other SN type. In Tab.~\ref{tab:global} we list the distance of the GRB from the brightest pixel in the H$\alpha$ map. Other studies take the brightest pixel in a broad-band filter (usually HST): \cite{Blanchard16} use whatever filter available, while \cite{Lyman17} only use the IR filter F160W, centered at 15,400 \AA{}. Some of the filters used include H$\alpha$ at the corresponding redshift, but others probe only continuum emission.

The average distance from the brightest spaxel is $1.25\pm0.93$\,kpc dominated by the large offset for GRB 050826, removing this host we get an average of $0.88\pm0.26$\,kpc. These values are well in agreement with distances reported in the literature: $1.4\pm0.8$\,kpc \citep{Bloom02}, $1.2\pm0.2$\,kpc \citep{Blanchard16} and $1.0\pm0.2$\,kpc \citep{Lyman17}.  Normalizing the distance from the brightest H$\alpha$ spaxel to the galaxy r$_\mathrm{80}$ gives an average normalized distance of $0.66\pm0.37$. \cite{Lyman17} found an average d$_\mathrm{GRB}$/r$_\mathrm{80}$ of only 0.3 (comprising hosts up to $z=2.7$), while a recent work comparing GRB and BL-Ic SN hosts without GRBs \citep{Japelj18} found  d$_\mathrm{GRB}$/r$_\mathrm{50}$ = 0.98 for GRB hosts (which partly comprises the sample presented here). The GRB location is also always inside the 80\% light radius of the host (see Tab. \ref{tab:global}). To test the hypothesis of a correlation between the SFR and the GRB location, we perform a Monte Carlo simulation, scaling the probability of a GRB being produced in a specific location with the H$\alpha$ flux. From this we obtain an expected average distance from the brightest pixel to the GRB location of 2.5$\pm$1.5\,kpc, which is larger than the distances that we measured. This indicates that GRBs do not occur in any random SF region, but are rather associated with massive, bright SF regions capable of producing the required massive progenitors.

\cite{Hammer06} proposed that GRBs might be runaway stars from extreme HII regions. GRB 980425 is 800\,pc from a bright region with WR lines, however, the GRB site itself is a smaller SF region \citep{Fynbo00}. For GRB\,020903, \cite{Hammer06} found a distance of 450\,pc from the brightest pixel of the SF region we named ``GRB region''. However, while none of the GRB sites are associated with the brightest HII region, they are always associated with some SF region. This could imply that a luminous SF region with a large number of massive stars is not the only determinig factor or that GRB progenitors are not necessarily the most massive stars.
 
The GRB is usually located in a region of the galaxy with low line width (see Fig. \ref{Fig:sample}). This is a common pattern in dwarf galaxies, where larger $\sigma$ values are found in between HII regions \citep[see e.g.][and references therein]{Cairos17}. In the three hosts where we resolve a broad component in different regions, the GRB region does clearly show an underlying broad component but it is usually not the region with the highest relative luminosity of the broad component. For the host of GRB 031203, the absolute flux of the broad component is highest in the GRB region, but the relative contribution is larger in the ``broad region''. In some cases, geometric effects might play a role in determining the actual observed strength of the broad component.

\subsection{The broad emission component and evidences for a star-burst wind}
All galaxies in our sample show an underlying broad component or high velocity emission. For GRB 030329 and GRB 050826 two separate components might be present in the blue and red wings (see Figs.\,\ref{fig:030329_Hafits} and \ref{fig:050826_Hafits}). Similar looking excess emission in H$\alpha$ has been observed in a sample of nearby extremely metal poor (XMP) dwarfs \citep{OlmoGarcia17} and explained with expanding shells that can be more or less symmetric and/or affected by differential dust extinction. Three galaxies of our sample also have multiple narrow components. Even galaxies that appear compact, however, can have multiple components in velocity due to the orientation of the galaxy, see e.g. the host of GRB 100418A \citep{deUgarte18b}.

The narrow and broad components are kinematically detached. While the narrow component follows the general velocity field the velocity of the broad component stays almost constant. The broad component is blueshifted, concentrated in the most intense SF region and more metal-rich. These results strongly hint to an outflow causing these components, since an inflow should have lower metallicity containing more pristine gas from the IGM.  

The possibly best example for an outflow is the host of GRB\,031203, where the emission spans 300--400 km\,s$^{-1}$ with possible additional components up to 700 km\,s$^{-1}$. This goes beyond the normal rotation field of a galaxy of this mass of $\sim$ 150 km\,s$^{-1}$. The broad component is blue-shifted and likely more metal-rich, adding to the outflow hypothesis. The broad component is strongest in the central parts and weaker in the Western ``tail'', implying that it might originate from the brightest SFR and is either blocked in some parts or not spherically symmetric. The excess emission in the red wing is stronger in the Western part but visible throughout the host. This might be an additional outflow component or shell or it could be connected to the radio emission from HI detected in \cite{Michalowski15}, claimed to be infalling, metal poor gas.  

For two hosts (GRB 060218 and GRB 030329) we can compare emission line kinematics from warm gas with absorption line kinematics from cold gas \citep{Wiersema07, Thoene07}. The absorption in GRB 030329 clearly supersedes the velocities from pure rotation, while for GRB 060218 the absorption lines of NaD match the components of the emitting gas. The two galaxies are similar in size and metallicity and both probably consist of a single large star-forming region. Galactic winds have been regularly detected also in NaD absorption \citep[see e.g.][]{Martin05, Veilleux05}, however, note that NaD has a different ionization energy than Mg I and II. One explanation could be that the outflow cones have small opening angles, hence for GRB 030329 the wind would point toward us while for GRB 060218 we see the galaxy outside of the wind cone. The fact that the host of GRB 030329 has hardly any velocity field while for GRB 060218 we do observe some rotation could point to a different inclination of these two galaxies.

 M82 shows a very similar narrow-broad profile, which can be traced out to large distances, and some regions at the base of the wind have double components associated with expanding shells \citep{Westmoquette09a, Westmoquette09b}. BCDs frequently show broad components, e.g. Haro 14 \citep{Cairos17}, Haro 11 \citep{Ostlin15}, NGC\,4449 \citep{kumari17}, UM448 \citep{James13}, Mrk996 \citep{James09} or NGC\,1569 \citep{Westmoquette07}. \cite{Westmoquette07} conclude that the broad component in the center of NGC\, 1569 originates from the interaction between a strong stellar wind and cold gas knots, producing a turbulent mixing layer on the surface which powers the observed starburst wind. Haro 11 \citep{Ostlin15} shows a triple component in  [S{\sc\,iii}], one with $\sigma \sim$~90\,km\,s$^{-1}$, but this component follows the stellar velocity field and might be a superposition of several unresolved lines. Outflows in absorption have been detected for the majority of starburst galaxies with fractions ranging from 75\% \citep{Chisholm15} to 90\% for face-on galaxies \citep{Heckman15}. Outflows in both absorption and emission have been detected in NGC 7552 \citep{Wood15}, a face-on spiral galaxy, showing blue-shifted emission components ($\sigma$ up to 300\,km\,s$^{-1}$) and blue-shifted absorption components up to 1000\,km\,s$^{-1}$.

At higher redshifts \cite{AmorinGPkin} find broad components in six GP galaxies at $z=0.1-0.3$ but with higher width than in our sample ($\sigma$ of 50--110 km\,s$^{-1}$, full width zero intensities $>$650\,km\,s$^{-1}$). They associate these components with stellar winds or supernova remnants and exclude turbulent mixing layers since they would only be observed in the Balmer lines but not the forbidden lines as it is also the case in our FLAMES sample. IFU spectra of z$\sim$0.2 GPs \citep{Lofthouse17} show evidence for a broad component with $\sigma$ $>$ 85~km\,s$^{-1}$. The extreme starburst in GPs together with a large number of WR stars could explain the larger velocities observed. GPs also show outflows in absorption and many are Lyman-$\alpha$ emitters, possible analogs to high redshift galaxies responsible for the escape of Ly$\alpha$ photons \citep[see e.g.][]{Henry15, Yang17}. Broad-narrow line profiles have been detected in massive star-forming galaxies at $z\sim 2$ in the SINS survey \citep[][]{Genzel11, Newman12, Davies19}. The blue-shifted broad profiles are associated with the brightest regions/clusters attributed to powerful winds in which the outflow rate can even supersede the SFR, quenching their own SF rather efficiently. 

Winds could also explain abundances differences in BDCs. In NGC\,4449, the central SF region is more metal poor than the outskirts of the galaxy, something also observed in other dwarf starbursts \citep{SanchezAlmeida15, Elmegreen16}. \cite{kumari17} propose an outflow of metal-rich gas acting stronger in the region with the highest SFR than in the outskirts, but also the inflow of metal-poor gas or pre-enriched gas giving rise to new SF regions after a merger are viable explanations. \cite{James09} found a large enhancement of N/O and nitrogen abundance in Mrk996 in the broad emission component while the narrow component shows a normal N/O ratio.  However, in UM448, WR features (albeit rather faint) are associated with a region showing a weaker broad component but an enhancement in N/H and N/O \citep{James13}. The authors speculate that WR stars alone cannot be responsible for the N/O enhancement and suggest an inflow of metal-poor gas leading to a decrease of the O abundance. Clearly, different galaxies are influenced by a varying interplay between in- and outflows, which are hard to study beyond the local Universe.

\subsection{Correlations between components and galaxy properties}\label{sect:discussionc}
\begin{figure*}[!ht]
\begin{center}
	\includegraphics[width=8cm]{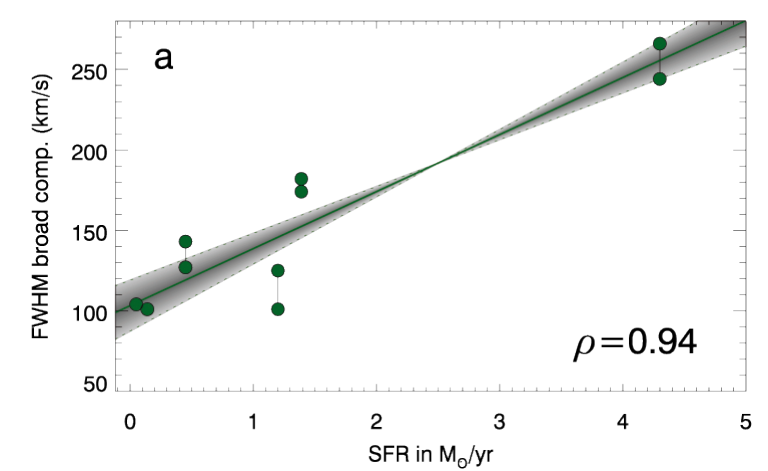}
	\includegraphics[width=8cm]{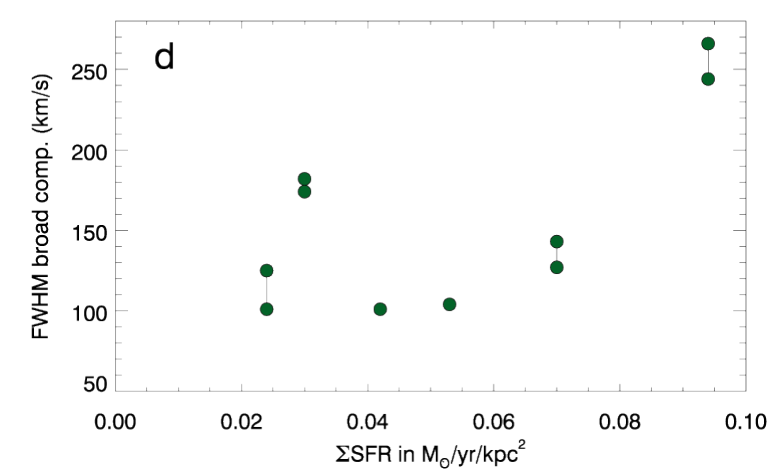}
	\includegraphics[width=8cm]{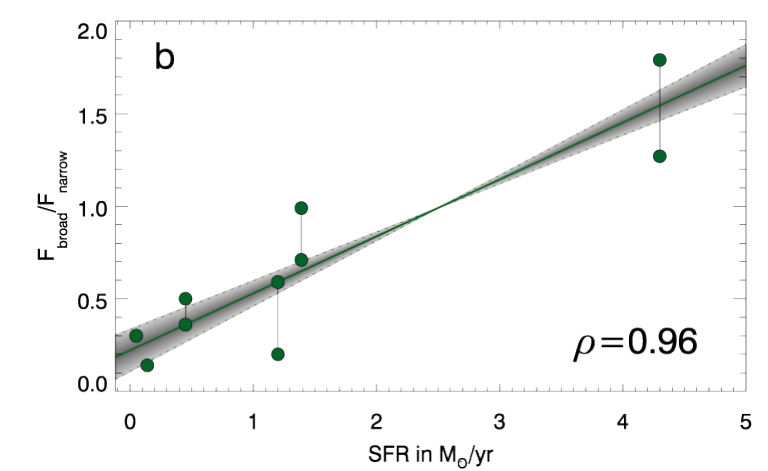}
	\includegraphics[width=8cm]{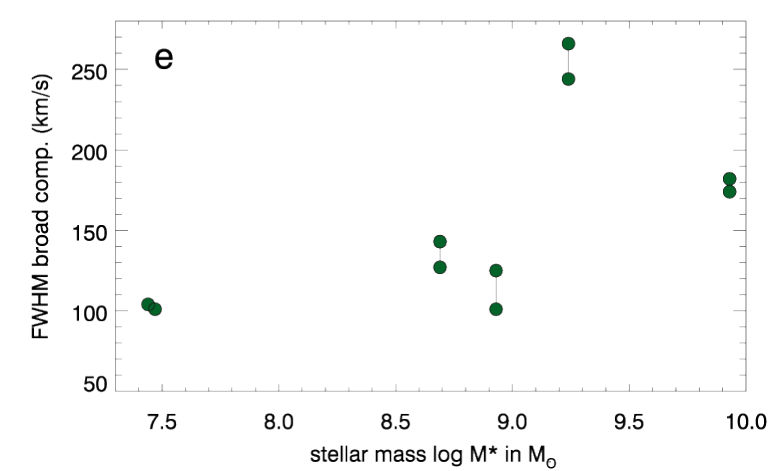}
	\includegraphics[width=8cm]{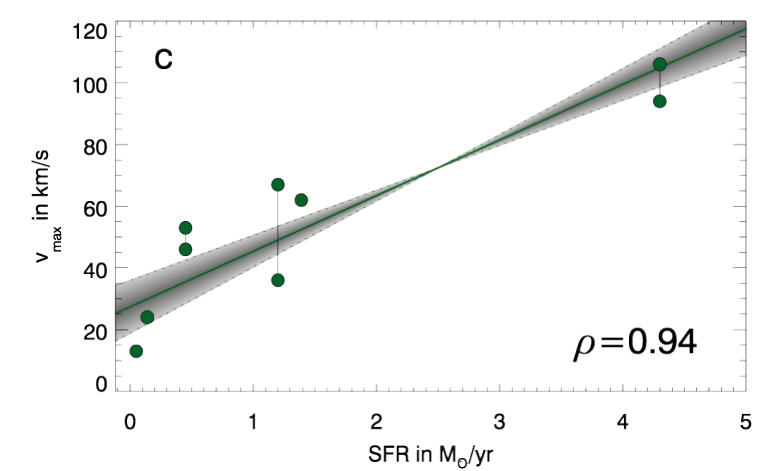}
	\includegraphics[width=8cm]{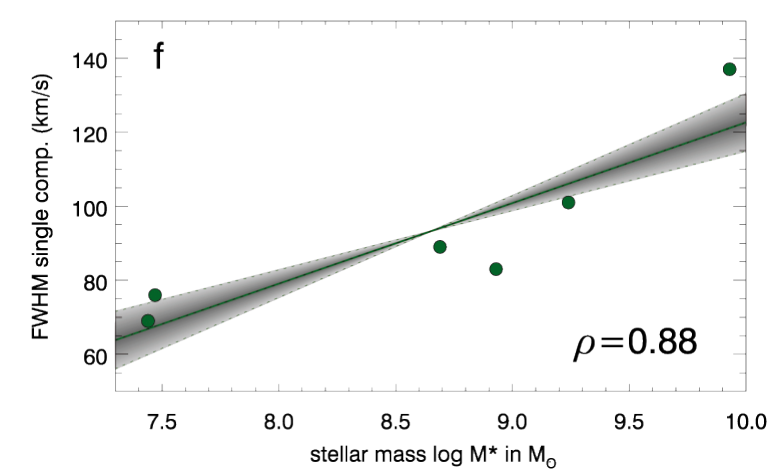}

    \caption{Panel A, B \& C: Correlations between SFR and FWHM of the broad emission component, F$_\mathrm{broad}$/F$_\mathrm{narrow}$ and V$_\mathrm{max}$ (see Sect. \ref{sect:discussionc}). For hosts with a range of values we plot the upper and lower limits. In panels with linear correlations, green lines are the linear fit, the shaded area is the error of the slope. In case of a range of values for the y-axis we take the average to fit the correlation. Panel D\&E: FWHM of the broad component vs. SFR density and M* for which there are no clear correlations. Panel F: A weak correlation is also found for M* and the width of a single Gaussian fit to the integrated galaxy spectra ($\sigma_\mathrm{int}$ in Tab.\,\ref{tab:globalkin}, here plotted as FWHM for consistency with the other plots), which corresponds to a stellar mass Tully Fisher relation. $\rho$ indicates the Pearson's coefficient for the different correlations.}
    \label{fig:correlations}
    \end{center}
\end{figure*}

We searched for possible correlations between the broad components and the properties of the host (stellar mass, luminosity, metallicity and SFR, see Tabs.~\ref{tab:global}, \ref{tab:globalkin}, and \ref{tab:kinfits} and Fig.~\ref{fig:correlations}). As an additional value we derived the maximum outflow velocity of the gas defined as
\begin{equation*}
V_\mathrm{max}\,=\,|\Delta\,v \mathrm{(broad-narrow)}-0.5\,\mathrm{FWHM}_\mathrm{broad}|
\end{equation*}
and listed in Tab.~\ref{tab:kinfits} \citep{Arribas14, Veilleux05}. For the hosts of GRBs\,020903, 050826 and 060218 we use $\Delta\,v$ between the bluest narrow and the broad component, for the host of GRB 030329 we use the blue shifted additional component. Note that in this section we use FWHM instead of $\sigma$. 

We only find correlations between the SFR and the broad component, namely with 1) the FWHM of the broad component (panel A), 2) the flux ratio between broad and narrow component(s) (panel B) and 3) the maximum velocity V$_\mathrm{max}$ (panel C). The only outlier from correlation 3) is GRB 020903 where the position of the broad component relative to the bluer narrow component shifts between negative and positive velocities. The Pearson's coefficient for all those correlations is good with values of 0.94, 0.96 and 0.94, while the (non) correlation between the FWHM and $\Sigma$SFR or the stellar mass yield coefficients of $<$0.7. Hosts with broader components, higher relative flux of the broad component, and higher V$_\mathrm{max}$ have also higher SFRs (see Fig. \ref{fig:correlations}). In the host with the highest SFR, GRB 031203, the flux of the broad component is higher than the one in the narrow component by a factor of 1.5, hence the contribution of the outflowing material is considerable compared to the emission from the bulk of the gas in the galaxy. We note that our sample might be too small to establish firm correlations, however, it is the largest currently available sample of GRB hosts which allows to search for those correlations.

We do not find a correlation between log M* and the width of the broad component (panel E) but there is some correlation between the FWHM fitting a single Gaussian and log M* (values listed in Tab.\,\ref{tab:globalkin} as $\sigma_\mathrm{int}$) with a Pearson's coefficient of 0.88 (panel F). The latter confirms the widely established ``stellar mass Tully-Fisher'' (sTF) M*-$\sigma$ relation \citep[see e.g.][]{Miller12}. The fact that the narrow component correlates with M*, but the broad component does not, confirms that the broad component is not related to the velocity field of the galaxy, as we have specifically shown for GRB 100316D (see Fig. \ref{fig:100316D_compflux}).

Several studies have tried to establish scaling relations between winds traced by absorption lines and SFR or stellar mass. They usually find the SFR, $\Sigma$SFR or SFR/M* to correlate with the outflow velocity (measured as $\sigma$ or V$_\mathrm{max}$) but with different slopes \citep[see e.g.][and references therein]{Erb12, Arribas14, Chisholm15, HeckmanBorthakur16}, but not the absolute SFR as it is the case for GRB hosts.  \cite{Heckman03} suggests that there is a minimum $\Sigma$SFR density of $\sim$0.1 M$_\odot$/y/kpc$^2$ needed to launch an outflow from a rotating disk, higher than the SFR densities observed in the FLAMES sample. Since the SFR density depends on the inclination angle, an incorrect inclination could change the value slightly. Extinction is not an issue as this would affect both $\Sigma$SFR and SFR. 

\cite{Tanner17} model a starburst wind and produce synthetic absorption lines. They only find a correlation with SFR and $\Sigma$SFR below a certain threshold, however the observed velocities can be much lower than the actual wind velocity. This also explains different slopes using absorption lines, since they trace SFRs above and below the threshold as well as different ions. \cite{Maseda14} do not find a correlation between M* and $\sigma$ (as measured from a single Gaussian fit) in a sample of extreme emission line galaxies (EELGs) at $z=1-2$, however, their mass range is rather small. Those galaxies have a $\sigma$ around 50 km\,s$^{-1}$ with some reaching up to 200~km\,s$^{-1}$ but they do not detect a broad component like \cite{AmorinGPkin} for GPs (which are simply EELGs at a certain redshift). \cite{Maseda14} conclude that the gas fraction must be higher for those galaxies in order to be stable, and part of the gas rapidly turns into stars before the starburst shuts down again. This process might be supported by winds or outflows. 

Comparing the FLAMES sample to a large sample using MANGA data \citep{Bruno19}, our galaxies have much higher relative strengths of the broad vs. narrow component but lower outflow speeds V$_\mathrm{max}$ and $\sigma$ of the broad component. Also, at the low masses of our sample galaxies, the detection rate of outflows should be rather low and not 100\% \citep{Bruno19}. We note, however, that the sample selection is different from the GRB hosts presented here, which might explain the differences. On the other hand, the low masses make it likely for the material to actually escape the galaxy. Using the dynamical mass of the galaxies derived in Sect. \ref{sect:discussion} (see Tab. \ref{tab:globalkin}), we can derive the escape velocity for gas at r=r$_\mathrm{80}$: 
\begin{equation*}
v_\mathrm{esc}\,=\sqrt{{{2M_\mathrm{dyn}\,G\,(1+ln(r_\mathrm{max}/r))}\over{3r}}}
\end{equation*}

 taking a ratio of r$_{\mathrm{max}}$/r = 10 where r$_{\mathrm{max}}$ is the maximum radius of an isothermal gravitational potential \citep{Arribas14}. For the hosts of  of GRB 031203 and GRB 100316D the outflowing gas might be able to escape the galaxy, for GRB 031203 V$_{\mathrm{max}}$ even superseeds V$_{\mathrm{esc}}$ by a factor of 2. The gas in the highest mass host, GRB 050826, does not reach escape velocity and, surprisingly, neither in the smallest host, GRB\,060218. \cite{Bruno19} and \cite{Arribas14} concluded that beyond log\,M$\sim$10.4, it is rare that the gas reaches escape velocity.

\subsection{The perils of low spatial resolution}\label{sect:resolved}
Our study shows the complications of kinematic studies with poorly resolved and/or longslit data. Integrated spectra usually show less components than the spectra of individual HII regions. For the host of GRB 100316D the integrated spectrum even displays an erroneous position of the broad component. The main problem is the fact that the broad component is almost constant in velocity, while the narrow component follows the rotation field. \cite{Moiseev12} analyzed the effect of decreasing resolution on IFU data in nearby starbursts. High flux values get smoothed out and the range of extreme $\sigma$ values decreases, however, the median $\sigma$ value remains constant. Regions with high $\sigma$ at the edge of HII regions blend together to larger structures of seemingly high $\sigma$.

One has to be especially careful with longslit data. If the slit is oriented along the velocity field, collapsing it to a single 1D spectrum can get broad components to be lost in the broadening of the narrow component due to the velocity shift. If the orientation of the galaxy is not known, it might lead to wrong conclusions on the presence/absence of different kinematic components. Another difficult case is hosts with double or multiple narrow components. Since the individual subcomponents can follow their own velocity fields, blending them together with the velocity change can mimic a velocity field of a normal rotating disk. However, this should be relatively easy to distinguish from the velocity field of a rotating disk.

\section{Conclusions}
This paper presents the first spatially resolved data of low-redshift long duration GRB hosts at high spectral resolution, able to distinguish different kinematical components. We study the kinematics using H$\alpha$ and include a limited study on abundances:

\begin{itemize}
\item Low redshift GRB hosts are often BCDs and dIrrs with properties similar to the general population of these galaxies.

\item Only two out of six hosts show a rotating disk. The two most compact hosts are dispersion dominated with very little rotation. The remaining two have a double narrow component which could point to systems in the process of merging.

\item The GRB is close to but not at the brightest region nor that with the highest $\sigma$. This might be an indication for kicks from a binary companion or some particular conditions in the SF region that give rise to the GRBs.

\item All galaxies have underlying broad components with a $\sigma$ of 50 -- 110 km\,s$^{-1}$ blueshifted compared to the main emission peak. We interpret these components as outflows due to winds from young stars and/or supernova explosions. For low mass hosts, the velocities of the wind (V$_\mathrm{max}$) supersede escape velocity and the gas might actually be capable of leaving the galaxy.

\item For GRB 030329, there is evidence for a metal-rich outflow from absorption lines, but GRB 060218, a very similar galaxy, does not show absorption lines at high velocities. Unfortunately, at low redshift we often lack absorption line spectra, while at high-$z$ those absorption components stretching over several 100 km\,s$^{-1}$ are detected frequently.

\item The strength of the broad component changes across the galaxy but the largest contribution or width is not necessarily associated with the region of highest SFR. The velocity field of the broad component does not follow the general rotation field of the galaxy.

\item The broad component seems to be more metal-rich, another indication for an outflow transporting enriched gas away from the galaxy via winds or SN explosions.

\item We find a correlation between the SFR of a host and 1) the width of the broad component, 2) the relative strength compared to the narrow component and 3) the maximum gas velocity V$_\mathrm{max}$, but no correlation with the stellar mass or the SFR density, the latter of which is usually correlated to the wind component in SF galaxies. In GRB hosts the broad component seems to be correlated with the total amount of current SF in the host. 

\item Detailed kinematics crucially depend on spectral and spatial resolution. Integration over too large regions can result in erroneous line profiles, especially if the integration is done across the velocity field or.
\end{itemize}

A larger sample with higher S/N and spectral resolution would be very much desired, but is difficult with current instrumentation. It would also be interesting to compare GRB hosts with those of SLSNe, which show even lower metallicities and higher SF rates \citep[see e.g.][]{Leloudas15, Schulze18}. A unique feature of GRBs is that, within a certain redshift range, we are able to probe both emission and absorption line kinematics, something which is very expensive to do e.g. for quasar intervening systems. The study of GRB host kinematics is still in its infancy, particularly in 3D, but can give us important indications on the processes of SF in starburst galaxies.

\begin{acknowledgements}

CT and AdUP acknowledge support from AYA2017-89384-P, CT and AdUP also from a Ram\'on y Cajal fellowships RyC-2012-09984 and RyC-2012-09975, LI from a Juan de la Cierva Integraci\'on fellowship IJCI-2016-30940. DAK acknowledges support from the Spanish National Research Project RTI2018-098104-J-I00 (GRBPhot). JFAF acknowledges support from the Spanish Ministerio de Ciencia, Innovación y Universidades through the grant PRE2018-086507. SDV acknowledges support from the French National Research Agency (ANR) under contract ANR-16-CE31-0003. LC is supported by YDUN grant DFF 4090-00079. 

Ground based observations were collected at the VLT under program 092.D-0389(A).
\end{acknowledgements}

\bibliographystyle{aa}
\bibliography{FLAMES_AA}




\appendix

\section{Information on the GRBs of the individual host galaxies}
\subsection{GRB 020903}
GRB\,020903 ($z=0.2506\pm0.0003$, \citealt{Bersier06}) was a soft X-ray flash (XRF) \citep{Sakamoto04} and belongs to the class of low-luminosity GRBs with an E$_{iso}$ of $1.1\times10^{49}$ ergs. The afterglow showed an initial rise during the first few hours \citep{Bersier06,Urata15}, however, data on the optical afterglow is sparse while the radio afterglow resembles regular afterglows in brightness and temporal behavior \citep{Soderberg04}. \cite{Soderberg05} and \cite{Bersier06} detected a SN with spectra similar to SN1998bw but 0.5--0.6\,mag fainter.

\subsection{GRB 030329}
GRB 030329 ($z=0.16867\pm0.00001$, \citealt{Thoene07}), detected by the HETE-2 satellite \citep{Vanderspek03}, represents the first spectroscopic association of a broad-line Type Ic SN with a cosmological long GRB, SN 2003dh \citep{Stanek03,Hjorth03, Matheson03}. It is still the GRB with the largest optical afterglow follow-up data set \citep[e.g.,][]{Lipkin04,Kann06} and its radio afterglow had been observed for over 13 years \citep{Mesler13,Peters19}. With an E$_{iso}$ $1.74\times10^{52}$ erg \citep{Kann10} it is also one of the most energetic GRBs at low redshift. 

\subsection{GRB 031203}
GRB\,031203 ($z=0.10536\pm0.00007$, \citealt{Margutti07}) was detected by INTEGRAL and is part of the nearby low-luminosity GRB population \citep{Sazonov04,Soderberg04}. It had a very weak afterglow and a bright but otherwise standard accompanying BL Type Ic SN 2003lw \citep{Malesani04,Gal-Yam04,Thomsen04,Cobb04}. The host galaxy has been extensively studied \citep[e.g.,][]{Prochaska04,Margutti07,Watson11,Guseva11,Symeonidis}.

\subsection{GRB 050826}
GRB 050826 ($z=0.296\pm0.001$, \citealt{Mirabal050826}) was a relatively low-luminosity GRB behind significant Galactic extinction. Its redshift was not reported until several months later \citep{HalpernGCN2}, and follow-up was very sparse \citep{Mirabal050826}. No SN follow-up has been reported.

\subsection{GRB 060218}
This very low redshift burst ($z=0.03342\pm0.00002$, \citealt{Pian06}) was an X-ray flash (XRF) with a very long duration of $\sim2100$ s, and is another low-luminosity GRB \citep{Soderberg06}. It may be associated with a magnetar \citep{Mazzali06,Toma07}and showed a peculiar afterglow with an additional thermal component in the X-rays and in the optical \citep{Campana06,Starling12}. interpreted as shock-breakout of the SN from the star \citep{Campana06,Waxman07,Li07}. The associated broead-line Type Ic SN 2006aj was extensively studied \citep{Pian06,Modjaz06,Mirabal06,Sollerman06,Cobb06,Ferrero06,Gorosabel06,Mazzali07,2007ApJ...663.1180K,2007ApJ...658L...5M}.

\subsection{GRB 100316D}
GRB 100316D was a very low redshift ($z=0.0592\pm0.0001$, \citealt{Bufano12}), long duration ($T_\mathrm{90}=1500$\,s), subluminous XRF with a soft prompt emission spectrum and a thermal component in X-rays \citep{Starling12}. Due to some confusion of the optical counterpart with bright star-forming regions in the host during the first days, there is very little information on the actual afterglow. However, the accompanying broad-lined Type Ic SN 2010dh was well-studied. \citep{Chornock10,Cano11,Olivares12,Bufano12}.

\section{Reanalysis of the metallicity of the host of GRB 050826}
\cite{Levesque10b} cite a supersolar metallicity based on long-slit spectra from LRIS/Keck. The FLAMES spectrum shows a low significance emission excess at the position of the [N{\sc\,ii}], however, at that metallicity, [N{\sc\,ii}]$\lambda$6585 should be strong. Assuming the same line width for H$\alpha$ and [N{\sc\,ii}], we derive a 3-$\sigma$ limit of [N{\sc\,ii}]/H$\alpha$\,$<$\,0.25, corresponding to a metallicity limit of 12+log(O/H)$\,<\,$8.45, in clear disagreement with the value in \cite{Levesque10b}. To investigate this further, we analyzed the original long-slit spectrum from \cite{Levesque10b} (D. Perley, priv. comm.) taken with a 1'' slit at an orientation of 343 degrees (almost N-S), hence covering most of the galaxy. In this spectrum we measure a ratio [N{\sc\,ii}]/H$\alpha$ of 0.2, in good agreement with our measurements in the FLAMES spectrum. The metallicity in \cite{Levesque10b} was derived using the R$_{23}$ parameter. However, the authors use the upper branch of this two-valued metallicity calibrator, which in this case is the wrong branch. Since the flux ratio from both datasets are consistent, the metallicity value stated in \cite{Levesque10b} is incorrect.

\section{SED fit with CIGALE}

Using photometric data from the literature (see Tables \ref{table:photometry} and \ref{table:031203phot}) we perform a Spectral Energy Distribution (SED) analysis of the host galaxy of the FLAMES sample with \texttt{CIGALE}\footnote{\url{https://cigale.lam.fr/}}  \citep{Burgarella05,Noll09,CIGALE} using its 2020 version. We consider a delayed star-formation history with an age for the main stellar population in the galaxies ranging from 500 Myr to 12 Gyr and an age range for the recent SF burst of 20--50 Myr to keep the computational effort at a reasonable degree. The delayed SFR function is based on \cite{Malek18}, which is again building on the considerations of \cite[eq. 3 in][]{Ciesla15}. This approach gives better estimates for the mass-weighted SFR than the case with a simple or double exponential SFR. We use an Initial Mass Function (IMF) as described in  \cite{Chabrier03} and a stellar population model from \cite{Bruzual03}, assuming a metallicity \textit{Z} with values of 0.008, 0.02 or 0.05.

Dust attenuation is treated differently for emission lines and continuum. To model the continuum extinction, we consider the modified \cite{2000ApJ...533..682C} attenuation law. The  slope $\delta$ for continuum extinction curve k$_\lambda$=A($\lambda$)/E(B--V) with k$_\lambda$ $\propto$ $\lambda^{\delta}$ was left to vary from --0.6 to 0.6 in steps of 0.2 (see \citet[][eq. 8]{CIGALE}). For the attenuation of the emission lines we use a MW \citep{Cardelli89}, LMC and SMC \citep{Pei92} extinction law with R$_\textnormal{V}=3.1$, 2.93 and 3.16 respectively. The color excess between emission lines and continuum attenuation, measured by the ratio f=E(B--V)$_\mathrm{cont.}$/E(B--V)$_\mathrm{em. lines}$ was determined to be 0.44, hence the younger population (causing the emission lines) has higher extinction than the older population. 
We also take into account the light from stars re-emitted in the IR by dust by using the \cite{Dale14} models. When we have both mid- and far-infrared photometry for some hosts, we adopt the slope for the dust mass heated by the radiation field to be $\alpha=2$, as found in \cite{Dale_2002}. When FIR data are not available, $\alpha_\textnormal{IR}$ can vary from 1 to 3, with steps of 0.5. No AGN component has been added to the analysis. For the final results we adopt the results fitting an SMC SED to the data.

%

\begin{table*}
\caption{Results for the SED fitting}
\label{table:resultsCIGALE}
\begin{tabular}{llccccccc}
\hline \hline
GRB & $z$ & $\log_{10}$SFR & $\log_{10}$M* & A$_V$ & E(B-V) & $Z$ & Reduced\ -\ \ $\chi^2$ & sSFR  \\
 & & \small{M$_{\odot}$ y$^{-1}$} & \small{M$_{\odot}$} & \small{mag} & \small{mag} & & & \small{Gyr$^{-1}$}\\
\hline
GRB100316D & 0.0591 & -0.16$_{-0.03}^{+0.02}$ & 9.39$_{-0.08}^{+0.07}$ & 0.08 $\pm$ 0.02 & 0.05 $\pm$ 0.00 & 0.0080 & 0.78 & 0.28 $\pm$ 0.05 \\
GRB060218 & 0.0331 & -0.91$_{-0.06}^{+0.05}$ & 7.40$_{-0.07}^{+0.06}$ & 1.09 $\pm$ 0.18 & 0.48 $\pm$ 0.11 & 0.0080 & 1.87 & 5.00 $\pm$ 0.98 \\
GRB050826 & 0.296 & 0.88$_{-0.30}^{+0.40}$ & 9.99$_{-0.65}^{+0.25}$ & 0.86 $\pm$ 0.67 & 0.47 $\pm$ 0.33 & 0.0080 & 0.03 & 0.77 $\pm$ 1.30 \\
GRB031203 & 0.1055 & 0.46$_{-0.05}^{+0.04}$ & 8.86$_{-0.12}^{+0.09}$ & 0.59 $\pm$ 0.08 & 0.33 $\pm$ 0.05 & 0.0080 & 3.73 & 3.94 $\pm$ 1.03 \\
GRB030329 & 0.169 & -0.80$_{-0.18}^{+0.13}$ & 7.70$_{-0.17}^{+0.12}$ & 0.33 $\pm$ 0.24 & 0.16 $\pm$ 0.10 & 0.0200 & 1.85 & 3.17 $\pm$ 1.52 \\
GRB020903 & 0.2506 & 0.25$_{-0.33}^{+0.18}$ & 8.94$_{-0.40}^{+0.21}$ & 0.25 $\pm$ 0.26 & 0.13 $\pm$ 0.14 & 0.0500 & 0.60 & 2.05 $\pm$ 1.64 \\
\hline 
GRB100316D & 0.0591 & -0.16$_{-0.03}^{+0.02}$ & 9.39$_{-0.08}^{+0.07}$ & 0.07 $\pm$ 0.02 & 0.05 $\pm$ 0.00 & 0.0080 & 0.78 & 0.28 $\pm$ 0.05 \\
GRB060218 & 0.0331 & -0.91$_{-0.06}^{+0.05}$ & 7.40$_{-0.07}^{+0.06}$ & 1.09 $\pm$ 0.18 & 0.49 $\pm$ 0.11 & 0.0080 & 1.93 & 4.97 $\pm$ 1.00 \\
GRB050826 & 0.296 & 0.86$_{-0.43}^{+0.37}$ & 9.98$_{-0.62}^{+0.25}$ & 0.85 $\pm$ 0.65 & 0.47 $\pm$ 0.32 & 0.0080 & 0.03 & 0.75 $\pm$ 1.18 \\
GRB031203 & 0.1055 & 0.46$_{-0.04}^{+0.04}$ & 8.86$_{-0.12}^{+0.09}$ & 0.58 $\pm$ 0.08 & 0.33 $\pm$ 0.05 & 0.0080 & 3.70 & 4.02 $\pm$ 1.05 \\
GRB030329 & 0.169 & -0.80$_{-0.18}^{+0.13}$ & 7.70$_{-0.17}^{+0.12}$ & 0.33 $\pm$ 0.24 & 0.16 $\pm$ 0.10 & 0.0200 & 1.85 & 3.17 $\pm$ 1.52 \\
GRB020903 & 0.2506 & 0.24$_{-0.30}^{+0.18}$ & 8.93$_{-0.41}^{+0.21}$ & 0.23 $\pm$ 0.24 & 0.13 $\pm$ 0.13 & 0.0500 & 0.61 & 2.00 $\pm$ 1.57 \\
\hline
GRB100316D & 0.0591 & -0.16$_{-0.03}^{+0.02}$ & 9.39$_{-0.08}^{+0.07}$ & 0.08 $\pm$ 0.02 & 0.05 $\pm$ 0.00 & 0.0080 & 0.78 & 0.28 $\pm$ 0.05 \\
GRB060218 & 0.0331 & -0.91$_{-0.06}^{+0.05}$ & 7.39$_{-0.07}^{+0.06}$ & 1.09 $\pm$ 0.18 & 0.48 $\pm$ 0.11 & 0.0080 & 1.88 & 5.00 $\pm$ 0.98 \\
GRB050826 & 0.296 & 0.88$_{-0.32}^{+0.39}$ & 9.99$_{-0.64}^{+0.25}$ & 0.86 $\pm$ 0.67 & 0.47 $\pm$ 0.33 & 0.0080 & 0.03 & 0.77 $\pm$ 1.29 \\
GRB031203 & 0.1055 & 0.46$_{-0.05}^{+0.04}$ & 8.86$_{-0.12}^{+0.09}$ & 0.59 $\pm$ 0.08 & 0.32 $\pm$ 0.05 & 0.0080 & 3.73 & 3.96 $\pm$ 1.04 \\
GRB030329 & 0.169 & -0.80$_{-0.18}^{+0.13}$ & 7.70$_{-0.17}^{+0.12}$ & 0.33 $\pm$ 0.24 & 0.16 $\pm$ 0.10 & 0.0200 & 1.85 & 3.17 $\pm$ 1.52 \\
GRB020903 & 0.2506 & 0.25$_{-0.33}^{+0.18}$ & 8.94$_{-0.40}^{+0.21}$ & 0.25 $\pm$ 0.26 & 0.13 $\pm$ 0.14 & 0.0500 & 0.60 & 2.05 $\pm$ 1.64 \\
\hline
\end{tabular}
\tablefoot{Fits were done with a Milky Way attenuation law, R$_V=3.1$ (top), Large Magellanic Cloud attenuation law, R$_V=3.16$ (middle) and Small Magellanic Cloud attenuation law, R$_V=2.93$}
\end{table*}

\begin{table}
\caption{Photometry used for the SED fitting for all hosts except the one of GRB\,031203.}            
\label{table:photometry}      
\centering                          
\scriptsize
\begin{tabular}{l c c c}        
\hline\hline                 
Band        & Instrument    & AB Magnitude & Reference  \\   
\hline  \hline                       
GRB\,020903& & &\\ \hline
I          &  CTIO/MosaicII          & 20.83 $\pm$ 0.10 & \cite{Bersier06} \\
R          &  CTIO/MosaicII          & 20.91 $\pm$ 0.10 & \cite{Bersier06} \\
F606W      &  HST/ACS                & 21.01 $\pm$ 0.05 & \cite{2007ApJ...657..367W} \\
V          &  Danish 1.54 Telescope  & 20.71 $\pm$ 0.10 & \cite{Bersier06} \\
B          &  Danish 1.54 Telescope  & 21.41 $\pm$ 0.10 & \cite{Bersier06} \\
\hline\hline                                    
GRB\,030329& & &\\ \hline
IRAC3  &  \textit{Spitzer}/IRAC  & $>$ 18.96        &  \cite{Svensson10} \\
IRAC1  &  \textit{Spitzer}/IRAC  & $>$ 22.59        &  \cite{Svensson10} \\
K'     &  CAHA/Omega2000         & $>$ 21.56        &  \cite{refId0} \\
H      &  CAHA/Omega2000         & 22.54 $\pm$ 0.24 &  \cite{refId0} \\
J      &  CAHA/Omega2000         & 22.40 $\pm$ 0.16 &  \cite{refId0} \\
F814W  &  HST/ACS                & 22.71 $\pm$ 0.05 &  \cite{2007ApJ...657..367W} \\
C4     &  CAHA/BUSCA             & 22.51 $\pm$ 0.04 &  \cite{refId0} \\
R      &  CAHA/MOSCA             & 22.81 $\pm$ 0.04 &  \cite{refId0} \\
C3     &  CAHA/BUSCA             & 22.76 $\pm$ 0.04 &  \cite{refId0} \\
F606W  &  HST/ACS                & 22.81 $\pm$ 0.05 &  \cite{2007ApJ...657..367W} \\
V      &  CAHA/MOSCA             & 22.78 $\pm$ 0.10 &  \cite{refId0} \\
B      &  CAHA/MOSCA             & 23.27 $\pm$ 0.07 &  \cite{refId0} \\
F435W  &  HST/ACS                & 24.11 $\pm$ 0.10 &  \cite{2007ApJ...657..367W} \\
C2     &  CAHA/BUSCA             & 22.80 $\pm$ 0.05 &  \cite{refId0} \\
U      &  CAHA/MOSCA             & 23.35 $\pm$ 0.10 &  \cite{refId0} \\ 
C1     &  CAHA/BUSCA             & 23.41 $\pm$ 0.03 &  \cite{refId0} \\ 
\hline\hline       
GRB\,050826& & &\\ \hline                      
I          &  Danish 1.54 Telescope          & 20.02 $\pm$ 0.20 & \cite{2007ApJ...662..294O} \\
R          &  Danish 1.54 Telescope          & 20.26 $\pm$ 0.10 & \cite{2007ApJ...662..294O} \\
V          &  Danish 1.54 Telescope          & 20.76 $\pm$ 0.20 & \cite{2007ApJ...662..294O} \\
B          &  Danish 1.54 Telescope          & 21.56 $\pm$ 0.30 & \cite{2007ApJ...662..294O} \\
\hline                                   
   & Not included in fit    &  &   \\    
R          &  Unknown          & 20.89 $\pm$ 0.06 & \cite{2007ApJ...662..294O} \\
V          &  Unknown          & 20.30 $\pm$ 0.05 & \cite{2007ApJ...662..294O} \\
\hline\hline       
GRB\,060218& & &\\ \hline                     
Ks                 &  VLT/ISAAC         & 19.77 $\pm$ 0.09 & \cite{2012ApJ...756..187H} \\
H                  &  2MASS             & 19.80 $\pm$ 0.22 & \cite{2007ApJ...663.1180K} \\
J                  &  2MASS             & 19.76 $\pm$ 0.16 & \cite{2007ApJ...663.1180K} \\
R$_\textnormal{c}$ &  SUBARU/FOCAS      & 20.03 $\pm$ 0.10 & \cite{2007ApJ...658L...5M} \\
V                  &  SUBARU/FOCAS      & 20.14 $\pm$ 0.10 & \cite{2007ApJ...658L...5M} \\
B                  &  SUBARU/FOCAS      & 20.02 $\pm$ 0.10 & \cite{2007ApJ...658L...5M} \\ 
F160W              &  HST/WFC3          & 19.70 $\pm$ 0.01 & \cite{Lyman17} \\
\hline
   &  Not included in fit   &  &   \\    
F160W       &  HST/WFC3          & 19.611 $\pm$ 0.003 & \cite{Blanchard16} \\
\hline\hline
GRB\,100316D& & &\\ \hline                        
WISE4     &  WISE                & 16.16 $\pm$ 0.51 & ALLWISE \\
WISE3     &  WISE                & 17.13 $\pm$ 0.17 & ALLWISE \\
WISE2     &  WISE                & 18.44 $\pm$ 0.05 & ALLWISE \\
WISE1     &  WISE                & 17.86 $\pm$ 0.03 & ALLWISE \\
z'        &  Gemini/GMOS-South   & 17.06 $\pm$ 0.06 & \cite{2011MNRAS.413..669C} \\
i'        &  Gemini/GMOS-South   & 17.10 $\pm$ 0.06 & \cite{2011MNRAS.413..669C} \\
r'        &  Gemini/GMOS-South   & 17.18 $\pm$ 0.07 & \cite{2011MNRAS.413..669C} \\
UVOT V    &  \textit{Swift}/UVOT & 17.38 $\pm$ 0.03 & \cite{2011MNRAS.411.2792S} \\
g'        &  Gemini/GMOS-South   & 17.46 $\pm$ 0.08 & \cite{2011MNRAS.413..669C} \\
UVOT B    &  \textit{Swift}/UVOT & 17.75 $\pm$ 0.03 & \cite{2011MNRAS.411.2792S} \\
UVOT U    &  \textit{Swift}/UVOT & 18.53 $\pm$ 0.03 & \cite{2011MNRAS.411.2792S} \\
UVOT UVW1 &  \textit{Swift}/UVOT & 18.91 $\pm$ 0.03 & \cite{2011MNRAS.411.2792S} \\
UVOT UVM2 &  \textit{Swift}/UVOT & 18.73 $\pm$ 0.03 & \cite{2011MNRAS.411.2792S} \\ 
UVOT UVW2 &  \textit{Swift}/UVOT & 18.82 $\pm$ 0.02 & \cite{2011MNRAS.411.2792S} \\ 
\hline                                   
\end{tabular}
\tablefoot{Magnitudes are in the AB system and have been corrected for Galactic extinction.}
\end{table}

\begin{table}
\caption{GRB 031203 host galaxy photometry.}             
\label{table:031203phot}      
\centering                          
\scriptsize
\begin{tabular}{c c c c}        
\hline\hline                 
Band        & Instrument     & Flux & Reference  \\    
            &                & mJy      &             \\ 
\hline                        
1.39 GHz    &  ATCA          & 0.191   $\pm$ 0.037  & \cite{Watson11} \\
3.45 GHz    &  APEX/LABOCA   &         $<$   0.012  & \cite{Watson11} \\
160$\mu$    &  Herschel/PACS & 47.000  $\pm$ 34.000 & \cite{Symeonidis} \\
100$\mu$    &  Herschel/PACS & 22.000  $\pm$ 16.000 & \cite{Symeonidis} \\
70$\mu$     &  Herschel/PACS & 39.000  $\pm$ 18.000 & \cite{Symeonidis} \\
WISE4       &  WISE          & 11.300  $\pm$ 1.000  & \cite{Symeonidis} \\
WISE3       &  WISE          & 1.740   $\pm$ 0.120  & \cite{Symeonidis} \\
WISE2       &  WISE          & 0.086   $\pm$ 0.009  & \cite{Symeonidis} \\
WISE1       &  WISE          & 0.110   $\pm$ 0.005  & \cite{Symeonidis} \\
I           &  VLT/FORS2     & 0.149   $\pm$ 0.005  & \cite{2006ApJ...645.1323M} \\
R           &  VLT/FORS2     & 0.108   $\pm$ 0.002  & \cite{2006ApJ...645.1323M} \\
V           &  VLT/FORS2     & 0.174   $\pm$ 0.008  & \cite{Margutti07} \\
B           &  VLT/FORS1     & 0.076   $\pm$ 0.004  & \cite{Margutti07} \\
U           &  VLT/FORS1     & 0.050   $\pm$ 0.008  & \cite{Margutti07} \\ 
\hline                                 
\end{tabular}
\tablefoot{For this host we list flux densities in mili-Jansky instead of magnitudes. Fluxes have been corrected for Galactic extinction.}
\end{table}

\begin{figure*}[!ht]
\begin{center}
	\includegraphics[width=\columnwidth]{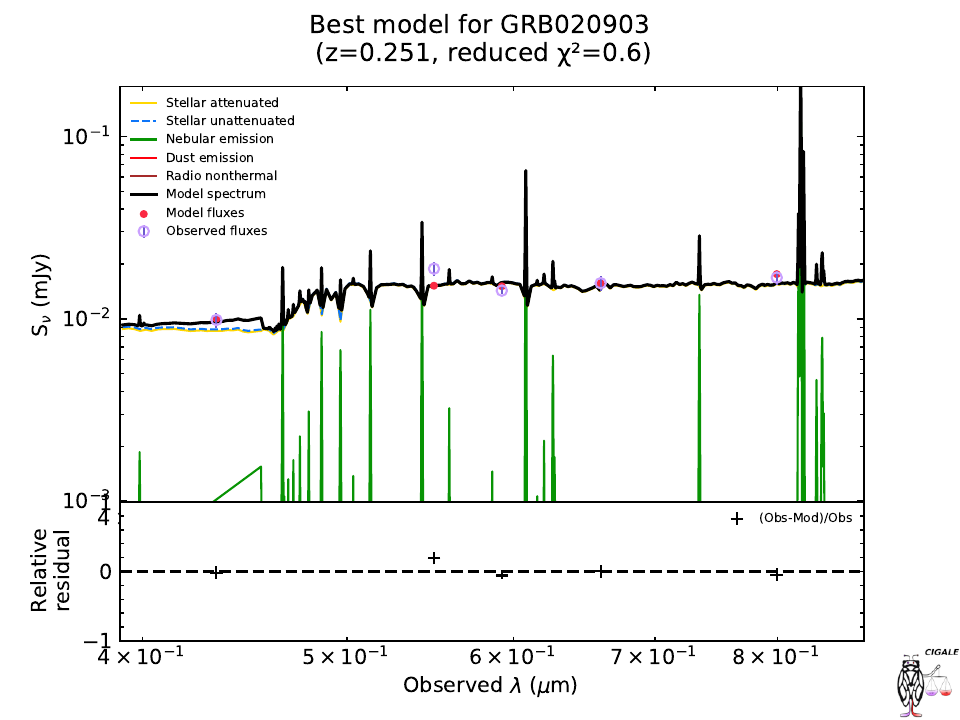}
	\includegraphics[width=\columnwidth]{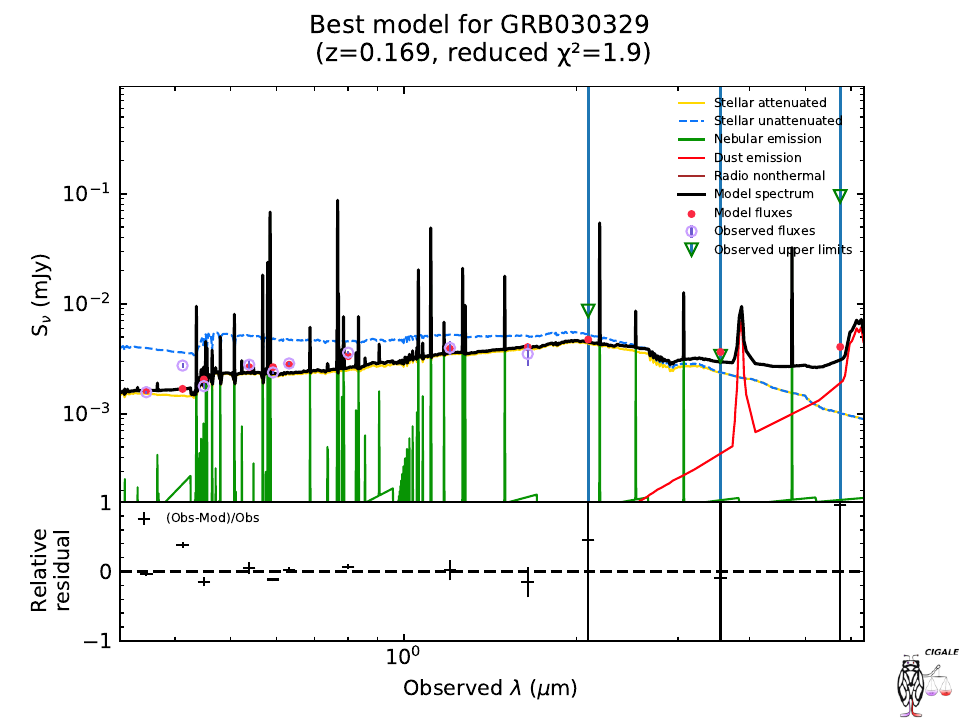}\\
	\includegraphics[width=\columnwidth]{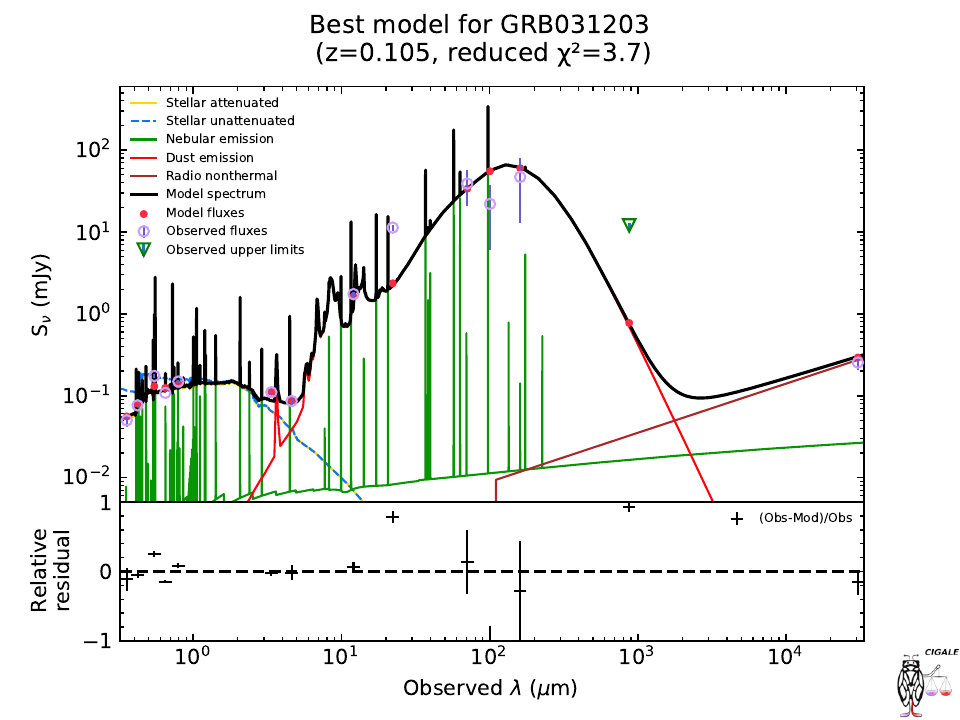}
	\includegraphics[width=\columnwidth]{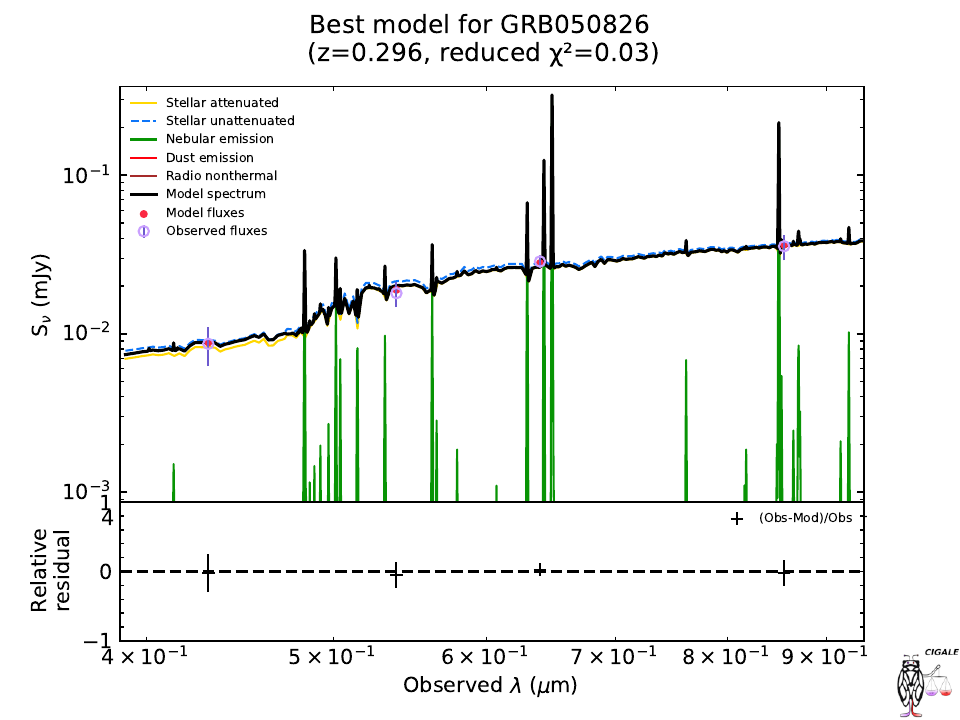}\\
	\includegraphics[width=\columnwidth]{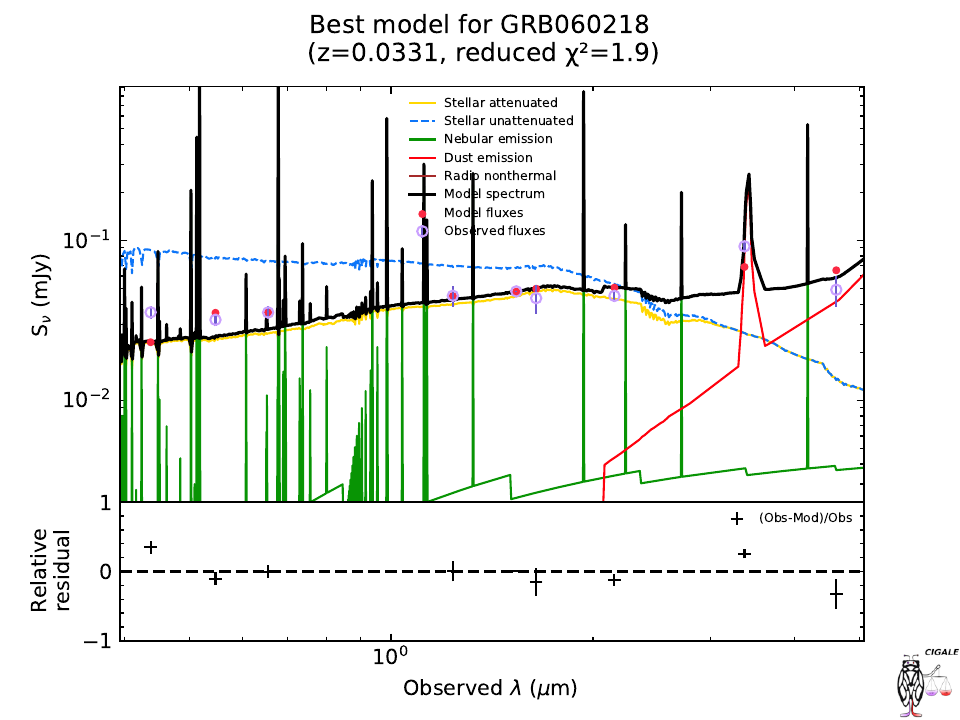}
	\includegraphics[width=\columnwidth]{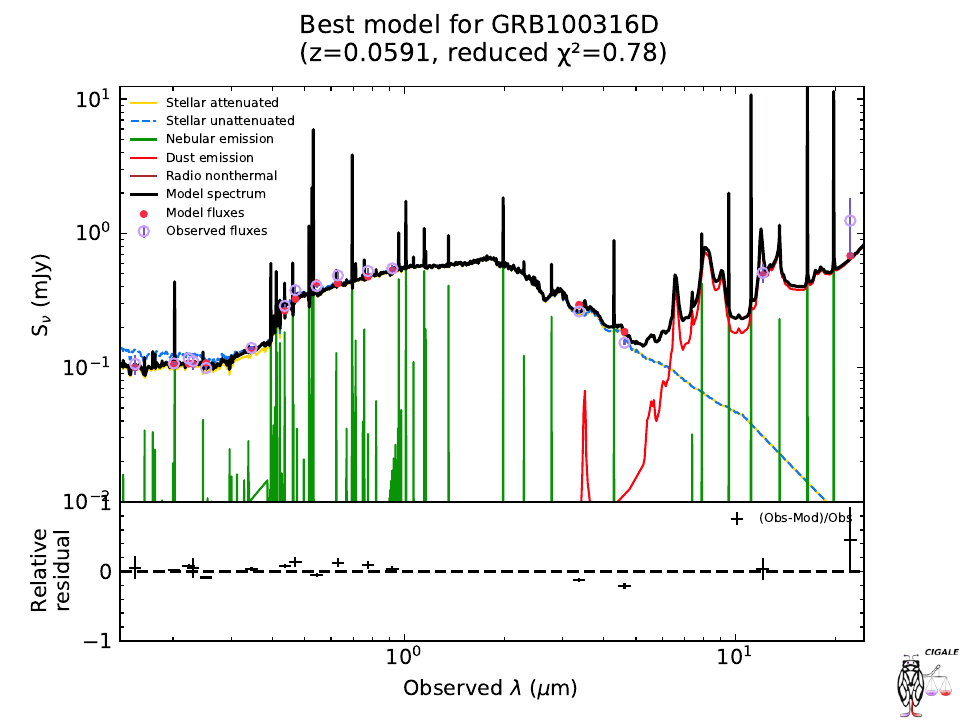}
	
    \caption{CIGALE SED fits of the hosts of GRB 020903, GRB 030329, GRB 031203, GRB 050826, GRB 060218 and GRB 100316D using the photometric data presented in Tab.\, \ref{table:photometry} and \ref{table:031203phot}.}
    \label{fig:cigale}
    \end{center}
\end{figure*}



\end{document}